\input{aipcheck}

\documentclass[
  ,draft            % use draft while you are working on the paper
,numberedheadings % uncomment this option for numbered sections
  ]
  {aipproc}

\layoutstyle{6x9}

\usepackage{color}
\usepackage{bm}
\newcommand{\be}{\begin{equation}}
\newcommand{\ee}{\end{equation}}
\newcommand{\ba}{\begin{eqnarray}}
\newcommand{\ea}{\end{eqnarray}}
\newcommand{\ff}[1]{{\bm #1}}
\newcommand{\tr}{\mbox{tr}}
\newcommand{\Tr}{\mbox{Tr}}
\newcommand{\refeq}[1]{Eq.\ (\ref{eq:#1})}
\newcommand{\labeq}[1]{\label{eq:#1}}

\newcommand{\bi}{\begin{itemize}}
\newcommand{\ei}{\end{itemize}}
\newcommand{\eq}[1]{\be #1 \ee}
\newcommand{\ca}[1]{{\cal #1}}
\newcommand{\ew}[1]{\langle #1 \rangle}
\newcommand{\ket}[1]{| #1 \rangle}
\newcommand{\bra}[1]{\langle #1 |}
\newcommand{\braket}[2]{\langle #1 | #2 \rangle}

\begin{document}

\title{Static and dynamic variational principles for strongly correlated electron systems}

\classification{71.10.-w, 71.10.Fd, 71.27.+a, 71.30.+h, 79.60.-i}
\keywords      {Correlated electrons, variational principles, static mean-field theory, dynamical mean-field theory}

\author{Michael Pottoff}{address={
I. Institut f\"ur Theoretische Physik, Universit\"at Hamburg, Jungiusstr. 9, 20355 Hamburg, Germany
}
}

\begin{abstract}
The equilibrium state of a system consisting of a large number of strongly interacting electrons can be characterized by its density operator. 
This gives a direct access to the ground-state energy or, at finite temperatures, to the free energy of the system as well as to other {\em static} physical quantities. 
Elementary excitations of the system, on the other hand, are described within the language of Green's functions, i.e.\ time- or frequency-dependent {\em dynamic} quantities which give a direct access to the linear response of the system subjected to a weak time-dependent external perturbation. 
A typical example is angle-revolved photoemission spectroscopy which is linked to the single-electron Green's function.
Since usually both, the static as well as the dynamic physical quantities, cannot be obtained exactly for lattice fermion models like the Hubbard model, one has to resort to approximations. 
Opposed to more ad hoc treatments, variational principles promise to provide consistent and controlled approximations.
Here, the Ritz principle and a generalized version of the Ritz principle at finite temperatures for the static case on the one hand and a dynamical variational principle for the single-electron Green's function or the self-energy on the other hand are introduced, discussed in detail and compared to each other to show up conceptual similarities and differences.
In particular, the construction recipe for non-perturbative dynamic approximations is taken over from the construction of static mean-field theory based on the generalized Ritz principle.
Within the two different frameworks, it is shown which types of approximations are accessible, and 
their respective weaknesses and strengths are worked out. 
Static Hartree-Fock theory as well as dynamical mean-field theory are found as the prototypical approximations.
\end{abstract}

\maketitle
\tableofcontents

%%%%%%%%%%%%%%%%%%%%%%%%%%%%%%%%%%%%%%%%%%%%
%% MAINMATTER
%%%%%%%%%%%%%%%%%%%%%%%%%%%%%%%%%%%%%%%%%%%%

\section{Motivation}

To understand the physics of systems consisting of a large number of interacting fermions constitutes one of the main and most important types of problems in physics. 
In condensed-matter physics many materials properties are governed, for example, by the interacting ``gas'' of valence electrons. 
From the theoretical perspective, the Coulomb interaction among the valence electrons must be considered as strong or at least of the same order of magnitude as compared to their kinetic energy for transition metals and their oxides, for example.
This implies that usual weak-coupling perturbation theory \cite{AGD64,FW71,NO88} does not apply. 
Density-functional theory (DFT) \cite{HK64,KS65,AvB85,JG89,Esc96} can be regarded as a standard technique in the field of electronic-structure calculations for condensed-matter systems. 
It provides an in principle exact approach which yields the electron density and the energy of the ground state.
In practice, however, it must be combined with approximations such as the famous local density approximation (LDA). 
While this DFT-LDA scheme has been proven to be extremely successful in predicting ground-state properties of a large class of materials, there are also several well-known shortcomings for so-called strongly correlated systems. 
These comprise many of 3d or 4f transition-metals and their oxides, for example. 
Another defect of the standard DFT consists in its inability to predict excited-state properties and the dynamic linear response. 
This is crucial, however, to make contact to experimental probes such as angle-resolved photoemission, for example. 
Interpretations of photoemission spectra are often based on the DFT-LDA band structure.
This lacks a fundamental justification and is is essentially equivalent to a Hartree-Fock-like picture of essentially independent electrons. 
The Hartree-Fock theory can be derived from a ``static'' variational principle where the ground-state energy or, at finite temperatures, the grand potential is minimized when expressed in a proper way as a functional of the pure or mixed state of the system, respectively. 
This is the Ritz variational principle.

Opposed to the static variational principle, however, there is a well-known ``dynamical'' variational principle which directly focuses on the one-electron excitation spectrum \cite{LW60}.
Here the grand potential is expressed as a functional of the one-electron Green's function or the self-energy and can be shown to be stationary at the respective physical quantity.
Similar to the density functional and similar to the Ritz principle, the dynamical variational principle is formally exact but needs additional approximations for a practical evaluation. 
Since long the approximations constructed in this way \cite{BK61,Bay62} have been perturbative as they are defined via partial resummations of diagrams where contributions at some finite order are missing. 
Hence, they are valid in the weak-coupling regime only.
The question arises whether it is possible to derive approximations from a dynamical variational principle which are non-perturbative and able to access the physics of strongly correlated electron systems where several interesting phenomena, like spontaneous magnetic order \cite{DDN98,BDN01}, correlation-driven metal-insulator transitions \cite{Mot49,Mot90,Geb97} or high-temperature superconductivity \cite{And87,OM00} emerge.

Rather than starting from the non-interacting Fermi gas as the reference point around which the perturbative expansion is developed, a local perspective appears to be more attractive for strongly correlated electron systems, in particular for prototypical lattice models with local interaction, such as the famous Hubbard model \cite{Hub63,Gut63,Kan63}.
The idea is that the local physics of a solid-state ion with a strong and due to screening effects essentially local Coulomb interaction is the more proper starting point for a systematic theory and that a self-consistent embedding of the ion in the lattice environment captures the main effects.
Since the invention of dynamical mean-field theory (DMFT) \cite{MV89,Jar92,GK92b,GKKR96,KV04}, a non-perturbative approximation with many attractive properties is available which just relies on this local perspective. 
The paradigmatic field of applications for the DMFT is the Mott-Hubbard metal-insulator transition \cite{Mot61,Geb97} which, at zero temperature, can be seen as prototypical quantum phase transition that is driven by the electron-electron interaction and cannot be captured by perturbative methods. 
``Mottness'', i.e.\ physical phenomena originating from a close parametric distance to the Mott transition or the Mott insulator, is also believed to be a possible key feature for an understanding of the many unusual and highly interesting properties of cuprate-based high-temperature superconductors.
This example shows that the DMFT, at least as a starting point for further methodical improvements, nowadays appears as an attractive approach to the electronic structure of unconventional materials.
In particular, there is the exciting perspective that, when combined with DFT-LDA, dynamical mean-field theory will ultimately be able to constitute a new standard for ab initio electronic-structure calculations with a high predictive power. 

The DMFT can be derived in an elegant way from the dynamical variational principle. 
The purpose of these lecture notes is to demonstrate how this is achieved and whether it is possible to derive similar or new approximations in the same way and to characterize the strengths and weak points of these ``dynamical'' approximations. 
The strategy to be pursued here is to first understand the formalism related to the static Ritz principle and to show up the differences but also the close analogies with the dynamic approach.

The notes are organized as follows:
The next section introduces the systems we are interested in and discusses on a general level the variational approach as such.
Sec.\ \ref{sec:3} then develops the static variational principle as a generalization of the Ritz principle. 
This is used in Sec.\ \ref{sec:4} to construct static mean-field theory.
To transfer the insight that has been gained from the static approach to the dynamic one, Sec.\ \ref{sec:5} introduces the concept of Green's functions and diagrammatic perturbation theory.
With this it becomes possible to define the central Luttinger-Ward functional and the self-energy functional which serve to set up the dynamical variational principle.
These points are discussed in Sec.\ \ref{sec:6}.
With the variational cluster approximation we give a standard example for a non-perturbative approximation constructed from the dynamical variational principle.
Consistency issues, symmetry breaking and the systematics of dynamical approximations are discussed in Sec.\ \ref{sec:8}. 
Sec.\ \ref{sec:9} particularly focuses on approximations related to dynamical mean-field theory.
A summary and the conclusions are given in Sec.\ \ref{sec:10}.

Secs.\ \ref{sec:2} -- \ref{sec:5} are written on a standard textbook level and can be understood with basic knowledge in many-body theory. 
The contents of Secs.\ \ref{sec:6} -- \ref{sec:9} is basically taken from Ref.\ \cite{Pot11a} but include some extensions and changes necessary for a self-contained presentation and to make the topic more accessible to the less experienced reader.

\section{Models and variational methods}
\label{sec:2}

We consider a system of electrons in thermodynamical equilibrium at temperature $T$ and chemical potential $\mu$. 
The Hamiltonian of the system $H = H(\ff t, \ff U) = H_0(\ff t) + H_{1}(\ff U)$ consists of a non-interacting part specified by one-particle parameters $\ff t$ and an interaction part with interaction parameters $\ff U$: 
\ba
  H_0(\ff t) &=& \sum_{\alpha\beta} t_{\alpha\beta} \:
  c^\dagger_\alpha c_\beta \; ,
\nonumber \\  
  H_{1}(\ff U) &=& \frac{1}{2} 
  \sum_{\alpha\beta\gamma\delta} U_{\alpha\beta\delta\gamma} \:
  c^\dagger_\alpha c^\dagger_\beta c_\gamma c_\delta \: .
\labeq{mod}
\ea
The index $\alpha$ refers to an arbitrary set of quantum numbers labeling an orthonormal basis of one-particle states $|\alpha\rangle$. 
As is apparent from the form of $H$, the total particle number $N = \sum_\alpha n_\alpha$ with $n_\alpha = c_\alpha^\dagger c_\alpha$ is conserved. 
$\ff t$ and $\ff U$ refer to the set of hopping matrix elements and interaction parameters and are formally given by:
\ba
  t_{\alpha\beta} &=& \bra{\alpha} \left( \frac{\ff p^2}{2m} +  V(\ff r) \right) \ket{\beta}
   \; ,
\nonumber \\  
  U_{\alpha\beta\delta\gamma} &=&
  {}^{(1)}\bra{\alpha}
  {}^{(2)}\bra{\beta}
  \frac{\mbox{const.}}{|\ff r^{(1)} - \ff r^{(2)}|}
  \ket{\gamma}^{(1)}
  \ket{\delta}^{(2)}
\ea
where $\ff p^2/2m$ is the electron's kinetic energy, $V(\ff r)$ the external potential and $\mbox{const.} / |\ff r^{(1)} - \ff r^{(2)}|$ the electrostatic Coulomb interaction between two electrons ``1'' and ``2''.
One has to be aware, however, that in many contexts $\ff t$ and $\ff U$ are merely seen as model parameters or considered as effective parameters which in addition account for effects not included explicitly in the Hamiltonian, such as metallic screening, for example.

The Hamiltonian describes the most general two-particle interaction. 
To give examples and to apply the techniques to be discussed below to a more concrete situation, it is sometimes helpful to focus on a less general model. 
The famous Hubbard model \cite{Hub63,Gut63,Kan63},
\be
  H = \sum_{ij\sigma} t_{ij} c^\dagger_{i\sigma}c_{j\sigma} + \frac{U}{2} \sum_{i\sigma} n_{i\sigma} n_{i-\sigma} \: ,
\ee
is a prototypical model for a system of strongly correlated electrons.
Here, electrons are assumed to hop over the sites of an infinitely extended lattice with a single spin-degenerate atomic orbital per lattice site: $\alpha= (i\sigma)$. 
The hopping integrals are assumed to be diagonal with respect to the spin index and to be spin-independent. 
Furthermore, the interaction is assumed to be strongly screened and to act only locally, i.e.\ two electrons must occupy the same lattice site $i$ to interact via the Hubbard-$U$.
Due to the Pauli principle, these electrons must then have opposite spin projections $\sigma=\uparrow,\downarrow$.

There are numerous and largely different many-body techniques for an approximate solution of the Hubbard model or for the more general model \refeq{mod}.
Here, we will concentrate on ground-state properties or properties of the system in thermal equilibrium and focus on two classes of approaches, namely techniques based on a
\bi
\item
``static'' variational principle $\delta \Omega[\rho] = 0$
\ei
as well as techniques based on a
\bi
\item
``dynamic'' variational principle $\delta \Omega[\ff \Sigma]=0$
\ei
which represent prototypical examples of different variants of variational principles.
These two classes of principles are different, and actually there is no (known) mapping between them.
On the other hand, there are a number of illuminating and apparent analogies which are worth to be discussed. 
Formally, the principles are exact. 
The static principle provides the exact state of the quantum system or, at finite temperature, the exact density matrix $\rho$ of the system in thermal equilibrium. 
The dynamical principle, on the other hand, yields the exact equilibrium self-energy $\ff \Sigma$ or Green's function of the system.
For all practical issues, it is clear, however, that approximations are necessary.

%*********************************************************************************
\begin{figure}[t]
  \centerline{\includegraphics[width=0.65\textwidth]{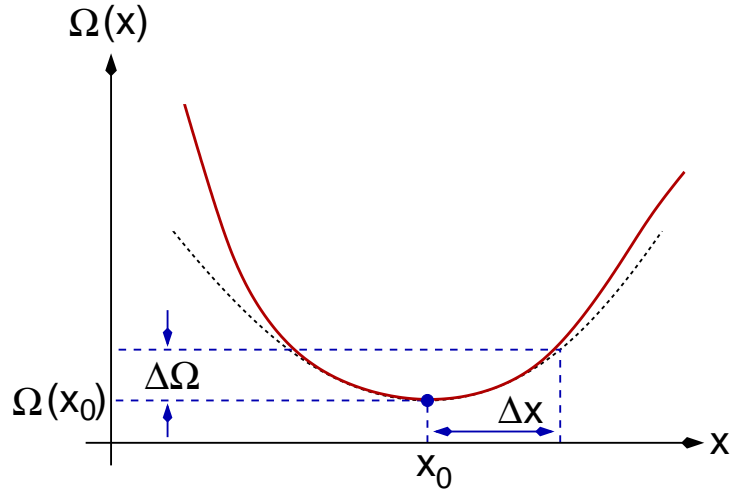}}
\caption{
The grand potential $\Omega$ is given as a function of some quantity $x$ (schematic).
At the physical value $x_0$ the grand potential is at a minimum $\Omega(x_0)$ corresponding to the physical value for $\Omega$.
Close to $x_0$ the function $\Omega(x)$ is quadratic to a good approximation (black dotted line).
Any approximate value $x_0 + \Delta x$ sufficiently close to $x_0$ therefore provides an approximate grand potential with an error $\Delta \Omega \propto \Delta x^2$ (blue dashed lines).
}
\label{fig:varprin}
\end{figure}
%*********************************************************************************
 
There are some obvious advantages of approximations constructed from a variational principle of the form $\delta \Omega(x) = 0$:
\bi
\item
The usual way to apply the variational principle is to propose some physically motivated form for the quantity of interest $x$ which may depend on a number of variational parameters $\ff \lambda = (\lambda_1, ..., \lambda_n)$. 
The optimal $x$ is then found by varying $\ff \lambda$ to find a set of parameters $\ff \lambda_0$ that satisfies $\partial \Omega (x(\ff \lambda_0)) / \partial \ff \lambda =0$.
This yields the approximation $x(\ff \lambda_0)$ to the exact $x_0$.
As there is not necessarily a small parameter involved, this way of constructing approximations is essentially {\em non-perturbative}. 
This also means, however, that the ansatz $x(\ff \lambda)$ has to be justified very carefully. 
\item
The variational procedure not only yields an approximation for $x_0$ but also for the grand potential $\Omega$.
As is obvious from Fig.\ \ref{fig:varprin}, if the approximate $x(\ff \lambda_0) = x_0 + \Delta x$ is sufficiently close to the exact or physical value $x_0$, i.e.\ if $\Delta x$ is sufficiently small, then the error in the grand potential is of {\em second order} only, $\Delta \Omega \propto \Delta x^2$.
\item
From the approximate grand potential one can derive, by differentiation with respect to parameters of the Hamiltonian, an in principle arbitrary set of physical quantities comprising thermal expectation values but also time-dependent correlation functions via higher-order derivatives.
As a rule of thumb, the higher the derivative the more accurate must be the approximate grand potential to get reliable estimates.
The fact that an approximate but explicit form for a thermodynamical potential is available, ensures that all quantities are derived {\em consistently}. 
E.g.\ the thermodynamical Maxwell's relations are fulfilled by construction.
\item
An approximation based on a variational principle can in most cases be generalized {\em systematically}. 
One simply has to allow for more variational parameters in the ansatz $x=x(\ff \lambda)$.
There is a clear tradeoff between the accuracy of the approximation on the one hand and the necessary computational effort to evaluate the resulting Euler equation $\partial \Omega (x(\ff \lambda_0)) / \partial \ff \lambda =0$ on the other hand when increasing the parameter space.
\item
If the grand potential is at a (global) minimum for the physical value $x_0$, then any approximation yields an {\em upper bound} to the physical $\Omega(x_0)$. 
This is an extremely helpful property since it allows to judge on the relative quality of an approximation (i.e.\ as compared to another one).
However, not all variational principles are minimum principles since usually $x$ is a multicomponent quantity. 
Then, $\delta \Omega(x_0)=0$ merely means that the grand potential is stationary at $x_0$ but it is not necessarily at a minimum (or maximum). 
As will be seen below, the static principle is a minimum principle while the dynamical variational principle is not.
\ei

\section{Static variational principle}
\label{sec:3}

To derive the static (generalized Ritz) variational principle, we will first compute the static response of an observable to a small static perturbation.
This will be used to prove the concavity of the grand potential which is necessary to derive the desired minimum principle. 

\subsection{Static response}

The grand potential of the system with Hamiltonian $H(\ff t, \ff U)$ at temperature $T$ and chemical potential $\mu$ is given by $\Omega_{\ff t,\ff U} = - T \ln Z_{\ff t,\ff U}$ where
\be
  Z_{\ff t,\ff U} = \tr \, \rho_{\ff t,\ff U} 
\labeq{partf}
\ee
is the partition function and 
\be
  \rho_{\ff t,\ff U} = \exp(-\beta(H(\ff t, \ff U) - \mu N)) 
\ee
the equilibrium density operator and $\beta=1/T$.
The dependence of the partition function (and of other quantities discussed below) on the parameters $\ff t$ and $\ff U$ is frequently made explicit through the subscripts.

Let $\lambda$ be a (one-particle or an interaction) parameter of the Hamiltonian that couples linearly to the observable $A$. 
We furthermore assume that the ``physical'' Hamiltonian is obtained for $\lambda = 1$, i.e.
$H(\ff t, \ff U) = H(\lambda=1)$ where $H(\lambda) \equiv H(\ff t, \ff U) - A + \lambda A \equiv H_0 + \lambda A$.
A straightforward calculation then yields 
\be
\frac{\partial \Omega}{\partial \lambda} = \langle A \rangle
\: .
\labeq{deri}
\ee
Note $H$ and $A$ do not necessarily commute and that the physical value of the expectation value is obtained by setting $\lambda=1$.

The computation of the second derivative is a bit more involved but also straightforward.
We have:
\be
\frac{\partial^2 \Omega}{\partial \lambda^2} 
=
\frac{\partial}{\partial \lambda} 
\left(
\frac{1}{Z} \tr \left(
A e^{-\beta( H(\lambda) - \mu N)}
\right)
\right)  \: .
\ee
Using $\partial Z/\partial \lambda = - \beta Z \,\ew{A}$, this yields
\be
\frac{\partial^2 \Omega}{\partial \lambda^2} 
=
\beta \ew{A}^2
+
\frac{1}{Z} \tr \left( A \frac{\partial}{\partial \lambda} 
  e^{-\beta (H(\lambda) - \mu {N})} \right)
  \: .
\ee
The derivative can be performed with the help of the Trotter decomposition:
\be
  \tr \, A \frac{\partial}{\partial \lambda} e^{-\beta (H_0 + \lambda  A- \mu {N})}
  =
  \tr \, A \frac{\partial}{\partial \lambda} \lim_{m\to \infty}
  \left( 
  e^{-\frac{\beta}{m} \lambda A }
  e^{-\frac{\beta}{m} (H_0 - \mu {N}) }
  \right)^m \: .
\ee
Writing $X=e^{-\frac{\beta}{m} \lambda A }e^{-\frac{\beta}{m} (H_0 - \mu {N}) }$ for short,
\be
\tr \, A \frac{\partial}{\partial \lambda} e^{-\beta (H_0 + \lambda  A- \mu {N})}
=
\tr \, A \frac{\partial}{\partial \lambda} \lim_{m\to \infty}
X^m
=
\lim_{m\to \infty}
\tr \, A 
\sum_{r=1}^m
X^{m-r} \frac{\partial X}{\partial \lambda} X^{r-1} \: .
\ee
With $\partial X/\partial \lambda = -\frac{\beta}{m} A X$ we get:
\be
\tr \, A \frac{\partial}{\partial \lambda} e^{-\beta (H_0 + \lambda  A- \mu {N})}
= - \tr
\lim_{m\to \infty} \, 
\sum_{r=1}^m
\frac{\beta}{m}
A  X^{m-r} A X^r \: .
\ee
In the continuum limit $m\to \infty$ we define $\tau = r \beta/m \in [0,\beta]$. 
Hence $d\tau = \beta/m$, and 
\be
\tr \, A \frac{\partial}{\partial \lambda} 
e^{-\beta(H-\mu N)} 
= 
- Z  
\int_0^\beta d\tau \langle
A(\tau) A(0)
\rangle
\ee
with the Heisenberg representation
\be
A(\tau) = e^{(H-\mu N)\tau} A  e^{- (H-\mu N)\tau}
\ee
for imaginary time $t=-i\tau$ where $0 < \tau < \beta$.
Collecting the results, we finally find:
\be
\frac{\partial^2 \Omega}{\partial \lambda^2} =
\beta \langle A \rangle^2 - \int_0^\beta d\tau \langle
A(\tau) A(0)
\rangle =
\frac{\partial \langle A \rangle}{\partial \lambda} \: .
\labeq{diss}
\ee
Physically, this is the response of the grand-canonical expectation value of the observable $A$ subjected to a small static external perturbation $d\lambda$.

Here the result can be used to show that the grand potential $\Omega$ is a concave function of $\lambda$:
\be
\frac{\partial^2 \Omega}{\partial \lambda^2} \le 0 \: .
\ee
This is seen as follows:
With $\Delta A = A-\ew{A}$ we have
\be
\frac{\partial^2 \Omega}{\partial \lambda^2} 
=
- \int_0^\beta d\tau \langle (A-\ew{A})(\tau) (A-\ew{A})(0)
\rangle 
\ee
Using the definition of the quantum-statistical average and $A=A^\dagger$, this implies
\be
\frac{\partial^2 \Omega}{\partial \lambda^2} 
=
- \int_0^\beta d\tau \frac{1}{Z} \tr \left(
e^{-\beta \ca H}
\Delta A (\tau/2)
\Delta A (\tau/2)^\dagger
\right) \le 0
\ee
Hence the grand potential is a concave function of any parameter $\lambda$ that linearly enters the Hamiltonian. 

\subsection{Generalized Ritz principle}

To set up the famous Ritz variational principle, we define
\be
  E_{\ff t, \ff U} [\ket{\Psi}] \equiv \frac{\bra{\Psi} H(\ff t, \ff U) \ket{\Psi}}{\braket{\Psi}{\Psi}} \: .
\ee
This represents the energy of the quantum systems as a functional of the state vector.
The functional parametrically depends on $\ff t$ and $\ff U$. 
The Ritz variational principle then states that the functional is at a (global) minimum for the ground state $\ket{\Psi_0(\ff t, \ff U)}$ of the system: 
\be
  E_{\ff t, \ff U} [\ket{\Psi_0(\ff t, \ff U)}] = \mbox{min.} \: 
\ee
The proof is straightforward and can be found in standard textbooks on quantum mechanics.

In the following we will generalize this principle to cover systems in thermal equilibrium with a heat bath at finite temperature $T$ and refer to this as the generalized Ritz principle. 
The classical version of the generalized principle goes back to Gibbs \cite{Gib48} and was lateron proven for quantum systems by von Neumann and Feynman \cite{vNeu55,Fey55,Mer65}.

Let us first define a functional which gives the grand potential of the system in terms of the density operator:
\be
\Omega_{\ff t, \ff U}[\rho] = \tr \Big( 
\rho ( H_{\ff t, \ff U} - \mu N + T \ln \rho )
\Big) \: .
\labeq{ritz}
\ee
Again, the functional parametrically depends on $\ff t$ and $\ff U$ as made explicit by the subscripts and on $T$ and $\mu$ (this dependence is suppressed in the notations).
The generalized Ritz principle then states that the grand potential is at a (global) minimum, 
\be
\Omega_{\ff t, \ff U}[\rho] = \mbox{min} \; ,
\ee
for the exact (the ``physical'') density operator of the system, i.e.\ for 
\be
\rho = \rho_{\ff t, \ff U} =
\frac{\exp(-\beta (H_{\ff t, \ff U} - \mu N))}
{\tr \exp(-\beta (H_{\ff t, \ff U} - \mu N))} \: ,
\labeq{rho0}
\ee
and that, if evaluated at the physical density operator, yields the physical value for the grand potential:
\be
\Omega_{\ff t, \ff U}[\rho_{\ff t, \ff U}] = \Omega_{\ff t, \ff U} =
- T \ln \tr \exp(-\beta (H_{\ff t, \ff U} - \mu N)) \: .
\ee

For the proof, we first note that the latter is satisfied immediately when inserting \refeq{rho0} into \refeq{ritz}.
Hence, it remains to show that $\Omega_{\ff t, \ff U}[\rho] \ge \Omega_{\ff t, \ff U}$ for ``arbitrary'' $\rho$.
The argument of the functional, however, should represent a physically meaningful density operator, i.e.\ $\rho$ shall be normalized ($\tr \rho = 1$), positive definite ($\rho \ge 0$) and Hermitian ($\rho=\rho^\dagger$).

To get a sufficiently general ansatz, we introduce the concept of a reference system.
This is an auxiliary system with a Hamiltonian 
\be
  H' = H(\ff t', \ff U') = H_0(\ff t') + \ff H_1(\ff U') 
\ee
that has the same structure as the Hamiltonian of the original model $H(\ff t, \ff U)$ but with different one-particle and interaction parameters.
The only purpose of the reference system is to span a space of trial density operators
\be
  \ca D \equiv \{ \rho_{\ff t',\ff U'} \: | \: \ff t', \ff U' \; \mbox{arbitrary} \}
\labeq{domain}
\ee
which are given as the exact density operators of the reference system when varying the parameters $\ff t'$ and $\ff U'$.
Hence, a trial $\rho$ is given by
\eq{
\rho = \rho_{\ff t' , \ff U'}
=
\frac{\exp(-\beta (H_{\ff t', \ff U'} - \mu N))}
{Z_{\ff t', \ff U'}} \: .
\labeq{ansatz}
}
\refeq{domain} defines the domain $\ca D$ of the functional \refeq{ritz}.
Note that the physical $\rho_{\ff t, \ff U} \in \ca D$.
Inserting \refeq{ansatz} in \refeq{ritz} we get:
\be
  \Omega_{\ff t, \ff U}[ \rho_{\ff t', \ff U'} ] 
  = 
  \ew{
  H_{\ff t, \ff U} - H_{\ff t' , \ff U'}}_{\ff t' , \ff U'} 
  + \Omega_{\ff t' , \ff U'} \: ,
\labeq{ins}
\ee
where the expectation value is done with respect to the reference system.

Now, consider the following the partition:
\be
H(\lambda)
\equiv
H_{\ff t', \ff U'} + \lambda (H_{\ff t, \ff U} - H_{\ff t', \ff U'}) \: .
\labeq{hl}
\ee
We have $H(0)=H_{\ff t', \ff U'}$ and $H(1) = H_{\ff t, \ff U}$ and with  
\be
\Omega(\lambda) \equiv - T \ln \tr \exp (-\beta (H(\lambda) - \mu N))
\ee
we get $\Omega(0)=\Omega_{\ff t', \ff U'}$ and $\Omega(1) = \Omega_{\ff t, \ff U}$.
The first term on the r.h.s.\ of \refeq{ins} represents the expectation value of a Hermitian operator that couples linearly via $\lambda$ to the Hamiltonian $H(\lambda)$, see \refeq{hl}. 
Using \refeq{deri} we can therefore immediately write \refeq{ins} in the form
\be
\Omega_{\ff t, \ff U}[ \rho_{\ff t', \ff U'} ] 
= 
\Omega(\lambda = 0)
+
\frac{\partial \Omega(\lambda)}{\partial \lambda} \Bigg|_{\lambda=0} \: .
\labeq{xx}
\ee
On the other hand, $\Omega(\lambda)$ is a concave function of $\lambda$, as has been shown in the preceding section.
Since any concave function is smaller than its linear approximation in some fixed point, e.g.\ in $\lambda=0$, we have:
\be
\Omega(0)
+
\frac{\partial \Omega(\lambda)}{\partial \lambda} \Bigg|_{\lambda=0}
\cdot \lambda
\ge 
\Omega(\lambda) \: .
\ee
Evaluating this relation for $\lambda=1$ and using \refeq{xx}, yields
\be
\Omega_{\ff t, \ff U}[ \rho_{\ff t', \ff U'} ] 
= 
\Omega(0)
+
\frac{\partial \Omega(\lambda)}{\partial \lambda} \Bigg|_{\lambda=0}
\ge
\Omega(1)
=
\Omega_{\ff t, \ff U} \: .
\ee
This proves the validity of the generalized Ritz principle.

\section{Using the Ritz principle to construct approximations}
\label{sec:4}

The standard application of the (generalized) Ritz principle is to construct the static mean-field approximation.
This represents the well-known Hartree-Fock approximation but generalized to systems at finite temperatures.

\subsection{Variational construction of static mean-field theory}

The general scheme to define variational approximations which can be evaluated in practice is to start from the variational principle $\delta \Omega_{\ff t,\ff U}[\rho]=0$ and to insert an ansatz for $\rho$ for which the functional can be evaluated exactly. 
To this end, one has to restrict the domain of the functional:
\be
  \rho \in \ca D' \subset \ca D \: .
\ee
Usually, this is necessary since the grand potential and the expectation value on the r.h.s.\ of \refeq{ins} are not available for interacting systems.
A restriction of the domain of the functional is equivalent with a restriction of the reference system, i.e.\ with a restricted set of parameters $\ff t'$ and $\ff U'$.
Any choice for $\ca D'$ results in a particular approximation.

Static mean-field theory emerges for the reference system 
\be
H' = H_{\ff t',0} = \sum_{\alpha\beta} t'_{\alpha\beta} c_\alpha^\dagger c_\beta
\ee
where the interaction term is dropped, $\ff U'=0$, but where all one-particle parameters $\ff t'$ are considered as variational parameters.
This is an auxiliary system of non-interacting electrons.
The corresponding restricted domain is:
\be
  \ca D' = \{ \rho_{\ff t',0} \: | \: \ff t' \; \mbox{arbitrary} \} \subset \ca D \: .
\ee
Hence, static mean-field theory aims at the optimal independent-electron density operator to describe an interacting system.
In the following we write 
\be
\rho_{\ff t'} \equiv \rho_{\ff t',\ff U=0} = \frac{1}{Z_{\ff t',0}} e^{-\beta(H' - \mu N)} 
\ee
and $\ew{\cdots}' \equiv \tr (\rho_{\ff t'} \cdots)$ for short.
Our goal is to determine the optimal set of variational parameters $\ff t'$ from the conditional equation
\be
\frac{\partial}{\partial t'_{\mu\nu}}
\Omega_{\ff t,\ff U}[\rho_{\ff t'}] = 0\: .
\labeq{cond}
\ee

To start with, we note
\be
\Omega_{\ff t,\ff U}[\rho_{\ff t'}] 
=
\tr \Big( 
\rho_{\ff t'} ( H_{\ff t, \ff U} - \mu N + T \ln \rho_{\ff t'} )
\Big) \: .
\ee
Inserting the trial density operator of the non-interacting reference system, we find:
\be
\Omega_{\ff t,\ff U}[\rho_{\ff t'}] 
=
\ew{H_{\ff t, \ff U} - \mu N}'
-
\ew{H_{\ff t',0} - \mu N}'
+ 
\Omega_{\ff t',0}
\ee
or 
\be
\Omega_{\ff t,\ff U}[\rho_{\ff t'}] 
=
\Big\langle
\sum_{\alpha\beta}
t_{\alpha\beta}
c_{\alpha}^\dagger
c_{\beta} 
+
\frac{1}{2}
\sum_{\alpha\beta\delta\gamma}
U_{\alpha\beta\delta\gamma} 
c_{\alpha}^\dagger
c_{\beta}^\dagger
c_\gamma
c_\delta 
-
\sum_{\alpha\beta}
t'_{\alpha\beta}
c_{\alpha}^\dagger
c_{\beta} 
\Big\rangle'
+
\Omega_{\ff t',0}
% \ew{c^\dagger_\mu c_\nu}'
\: .
\ee
The dependence on the variational parameters $t'_{\mu\nu}$ is twofold: 
There is an explicit dependence that is obvious in the third and the fourth term on the r.h.s.\ and there is an additional implicit dependence via the expectation value $\ew{\cdots}'$.
To calculate the derivative in \refeq{cond}, we first note that according to \refeq{deri}
$\partial \Omega_{\ff t',0} / \partial t'_{\mu\nu} = \ew{c^\dagger_\mu c_\nu}'$.
Furthermore, we define (see \refeq{diss}):
\be
K'_{\alpha\nu\mu\beta} 
= \frac{\partial \ew{c_\alpha^\dagger c_\beta}'}{\partial t'_{\mu\nu}}
= \frac{1}{T} \ew{c_\alpha^\dagger c_\beta}' \ew{c_\mu^\dagger c_\nu}'
- \int_0^{\beta} d\tau 
\ew{c_\alpha^\dagger(\tau) c_\beta (\tau)c_\mu^\dagger c_\nu}' \: .
\ee
Therewith, \refeq{cond} reads:
\be  
\sum_{\alpha\beta}
t_{\alpha\beta}
K'_{\alpha\nu\mu\beta} 
+
\frac{1}{2}
\sum_{\alpha\beta\gamma\delta}
U_{\alpha\beta\delta\gamma} 
\frac{\partial}{\partial t'_{\mu\nu}} 
\ew{c_{\alpha}^\dagger
c_{\beta}^\dagger
c_\gamma
c_\delta 
}'
-
\sum_{\alpha\beta}
t'_{\alpha\beta}
K'_{\alpha\nu\mu\beta} 
= 0 \:.
\ee

At this point we can make use of the fact that the reference system is given by a Hamiltonian that is bilinear in the creators and annihilators.
In this case Wick's theorem (see e.g.\ Ref.\ \cite{AGD64,FW71,NO88}) applies: 
Any $2N$-point correlation function consisting of $N$ creators and $N$ annihilators in an expectation value with respect to a bilinear Hamiltonian can be simplified and written as the sum over all different full contractions. 
Here, a full contraction is a distinct factorization of the $2N$-point correlation function into a product of $N$ two-point correlation functions.
Usually, Wick's theorem is formulated for time-ordered product of creators and annihilators, and time ordering produces, in the case of fermions, a minus sign for each transposition of these operators.
For the static expectation value encountered here, we only have to consider a creator to be ``later'' than an annihilator, to realize that in this sense the expectation value in the second term on the r.h.s.\ is already time ordered and to take care of the minus sign when ordering the resulting two-point correlation functions.
We find:
\be
\ew{c_{\alpha}^\dagger
c_{\beta}^\dagger
c_\gamma
c_\delta 
}'
=
\ew{c_{\alpha}^\dagger
c_\delta 
}'
\ew{
c_{\beta}^\dagger
c_\gamma
}'
-
\ew{c_{\alpha}^\dagger
c_\gamma 
}'
\ew{
c_{\beta}^\dagger
c_\delta
}' \: .
\ee
The result can also be derived in a more direct (but less elegant) way without using Wick's theorem, of course.
Using this and rearranging terms, we have:
\be 
\sum_{\alpha\beta\gamma\delta}
U_{\alpha\beta\delta\gamma} 
\ew{c_{\alpha}^\dagger
c_{\beta}^\dagger
c_\gamma
c_\delta 
}'
=
\sum_{\alpha\beta\gamma\delta}
\left(
U_{\alpha\beta\delta\gamma} 
-
U_{\alpha\beta\gamma\delta} 
\right) 
\ew{c_{\alpha}^\dagger
c_\delta 
}'
\ew{
c_{\beta}^\dagger
c_\gamma
}' \: .
\ee
We carry out the differentiation:
\ba
&&
\frac{\partial }{\partial t'_{\mu\nu}}
\sum_{\alpha\beta\gamma\delta}
U_{\alpha\beta\delta\gamma} 
\ew{c_{\alpha}^\dagger
c_{\beta}^\dagger
c_\gamma
c_\delta 
}'
\nonumber \\
&&=
\sum_{\alpha\beta\gamma\delta}
\left(
U_{\alpha\beta\delta\gamma} 
-
U_{\alpha\beta\gamma\delta} 
\right) 
\left(
\ew{
c_{\alpha}^\dagger
c_\delta
}'
K'_{\beta\nu\mu\gamma}
+
K'_{\alpha\nu\mu\delta}
\ew{
c_{\beta}^\dagger
c_\gamma
}'
\right)
\ea
and again rearrange terms to get:
\ba
&&\frac{\partial }{\partial t'_{\mu\nu}}
\sum_{\alpha\beta\gamma\delta}
U_{\alpha\beta\delta\gamma} 
\ew{c_{\alpha}^\dagger
c_{\beta}^\dagger
c_\gamma
c_\delta 
}'
\nonumber \\
&&=
\sum_{\alpha\beta\gamma\delta}
\left(
(
U_{\gamma\alpha\delta\beta} 
+
U_{\alpha\gamma\beta\delta} 
)
-
(
U_{\gamma\alpha\beta\delta} 
+
U_{\alpha\gamma\delta\beta} 
)
\right) 
\ew{
c_{\gamma}^\dagger
c_\delta
}'
K'_{\alpha\nu\mu\beta}
\ea
and thus
\be
\frac{\partial }{\partial t'_{\mu\nu}}
\sum_{\alpha\beta\gamma\delta}
U_{\alpha\beta\delta\gamma} 
\ew{c_{\alpha}^\dagger
c_{\beta}^\dagger
c_\gamma
c_\delta 
}'
=
2
\sum_{\alpha\beta\gamma\delta}
\left(
U_{\alpha\gamma\beta\delta} 
-
U_{\gamma\alpha\beta\delta} 
\right) 
\ew{
c_{\gamma}^\dagger
c_\delta
}'
K'_{\alpha\nu\mu\beta}
\:,
\ee
where we made use of $U_{\alpha\beta\delta\gamma} = U_{\beta\alpha\gamma\delta}$.
Collecting the results, we have
\be  
\sum_{\alpha\beta}
t_{\alpha\beta}
K'_{\alpha\nu\mu\beta} 
-
\sum_{\alpha\beta}
t'_{\alpha\beta}
K'_{\alpha\nu\mu\beta} 
+
\sum_{\alpha\beta\gamma\delta}
\left(
U_{\alpha\gamma\beta\delta} 
-
U_{\gamma\alpha\beta\delta} 
\right) 
\ew{
c_{\gamma}^\dagger
c_\delta
}'
K'_{\alpha\nu\mu\beta}
= 0
\ee
or
\be  
\sum_{\alpha\beta}
\left(
t_{\alpha\beta}
-
t'_{\alpha\beta}
+
\sum_{\gamma\delta}
\left(
U_{\alpha\gamma\beta\delta} 
-
U_{\gamma\alpha\beta\delta} 
\right) 
\ew{
c_{\gamma}^\dagger
c_\delta
}'
\right)
K'_{\alpha\nu\mu\beta} = 0\: .
\ee
Assuming that $K$ can be inverted, this implies
\be
t'_{\alpha\beta}
=  
t_{\alpha\beta}
+
\sum_{\gamma\delta}
\left(
U_{\alpha\gamma\beta\delta} 
-
U_{\gamma\alpha\beta\delta} 
\right) 
\ew{
c_{\gamma}^\dagger
c_\delta
}' \: .
\labeq{tprim}
\ee
Hence, the optimal one-particle Hamiltonian of the reference system reads:
\be
H'
=
\sum_{\alpha\beta}
\left( 
t_{\alpha\beta}
+
\Sigma^{\rm (HF)}_{\alpha\beta} 
\right)
c_\alpha^\dagger c_\beta
\ee
where
\be
\Sigma^{\rm (HF)}_{\alpha\beta} 
=
\sum_{\gamma\delta}
\left(
U_{\alpha\gamma\beta\delta} 
-
U_{\gamma\alpha\beta\delta} 
\right) 
\ew{
c_{\gamma}^\dagger
c_\delta
}'
\labeq{hfsig}
\ee
is the (frequency-independent) Hartree-Fock self-energy.
Note that the self-energy has to be determined self-consistently: 
Starting with a guess for $\ff \Sigma^{\rm (HF)}$, we can fix the reference system's Hamiltonian $H'$.
The two-point correlation function of the reference system $\ew{c_\alpha^\dagger c_\beta}'$ is then easily calculated by a unitary transformation of the one-particle basis set $\alpha \mapsto k$ such that the correlation function becomes diagonal, $\ew{c_k^\dagger c_{k'}}' \propto \delta_{kk'}$, and by using Fermi gas theory to get 
$\ew{c_k^\dagger c_k}'$ from the Fermi-Dirac distribution and, finally, by back-transformation to find  $\ew{c_\alpha^\dagger c_\beta}'$. 
With this, a new update of the Hartree-Fock self-energy is obtained from \refeq{hfsig}.

The first term in \refeq{hfsig} is the so-called Hartree potential. 
It can be interpreted classically as the electrostatic potential of the charge density distribution resulting from the $N$ electrons of the system. 
Opposed to the first term, the second one is spatially non-local if written in real-space representation. 
This is the Fock potential produced by the $N$ electrons and has no classical analogue. 
Note that there is no self-interaction of an electron with the potential generated by itself: 
Within the real-space representation, the corresponding Hartree and Fock terms are seen to cancel each other exactly.
 
\subsection{Grand potential within static mean-field theory}

The final task is to compute the grand potential for the optimal (Hartree-Fock) density operator, i.e.
\be
\Omega_{\ff t,\ff U}^{\rm (HF)}
=
\Omega_{\ff t,\ff U}[\rho_{\ff t'}]
=
\tr \Big( 
\rho_{\ff t'} ( H_{\ff t, \ff U} - \mu \hat{N} + T \ln \rho_{\ff t'} )
\Big)
\ee
where (the self-consistent) $\ff t'$ is taken from \refeq{tprim}.
We find:
\be
\Omega_{\ff t,\ff U}^{\rm (HF)}
= 
\Omega_{\ff t',0}
+
\Big\langle
\sum_{\alpha\beta}
t_{\alpha\beta}
c_{\alpha}^\dagger
c_{\beta} 
+
\frac{1}{2}
\sum_{\alpha\beta\delta\gamma}
U_{\alpha\beta\delta\gamma} 
c_{\alpha}^\dagger
c_{\beta}^\dagger
c_\gamma
c_\delta 
-
\sum_{\alpha\beta}
t'_{\alpha\beta}
c_{\alpha}^\dagger
c_{\beta} 
\Big\rangle' \: .
\labeq{hff}
\ee
Using Wick's theorem,
\ba
\Omega_{\ff t,\ff U}^{\rm (HF)}
&=&
\Omega_{\ff t',0}
+
\sum_{\alpha\beta}
t_{\alpha\beta}
\ew{c_{\alpha}^\dagger
c_{\beta} 
}'
\nonumber \\
&+&
\frac{1}{2}
\sum_{\alpha\beta\gamma\delta}
\left(
U_{\alpha\beta\delta\gamma} 
-
U_{\alpha\beta\gamma\delta} 
\right) 
\ew{c_{\alpha}^\dagger
c_\delta 
}'
\ew{
c_{\beta}^\dagger
c_\gamma
}'
-
\sum_{\alpha\beta}
t'_{\alpha\beta}
\ew{c_{\alpha}^\dagger
c_{\beta} 
}'
\ea
and inserting the optimal $t'_{\alpha\beta}
=  
t_{\alpha\beta}
+
\sum_{\gamma\delta}
\left(
U_{\alpha\gamma\beta\delta} 
-
U_{\gamma\alpha\beta\delta} 
\right) 
\ew{
c_{\gamma}^\dagger
c_\delta
}'
$, we arrive at:
\ba
\Omega_{\ff t,\ff U}^{\rm (HF)}
= 
\Omega_{\ff t',0}
&+&
\frac{1}{2}
\sum_{\alpha\beta\gamma\delta}
\left(
U_{\alpha\beta\delta\gamma} 
-
U_{\alpha\beta\gamma\delta} 
\right) 
\ew{c_{\alpha}^\dagger
c_\delta 
}'
\ew{
c_{\beta}^\dagger
c_\gamma
}'
\nonumber \\
&-&
\sum_{\alpha\beta\gamma\delta}
 \left(
U_{\alpha\gamma\beta\delta} 
-
U_{\gamma\alpha\beta\delta} 
\right) 
\ew{
c_{\gamma}^\dagger
c_\delta
}'
\ew{c_{\alpha}^\dagger
c_{\beta} 
}' \: .
\ea
With the substitution $(\alpha\beta\gamma) \to (\beta\gamma\alpha)$ in the second term, this yields
\be
\Omega_{\ff t,\ff U}^{\rm (HF)}
= 
\Omega_{\ff t',0}
-
\frac{1}{2}
\sum_{\alpha\beta\gamma\delta}
\left(
U_{\alpha\beta\delta\gamma} 
-
U_{\alpha\beta\gamma\delta} 
\right) 
\ew{c_{\alpha}^\dagger
c_\delta 
}'
\ew{
c_{\beta}^\dagger
c_\gamma
}'
=
\Omega_{\ff t,\ff U}^{\rm (HF)}
-
\frac{1}{2}
\sum_{\alpha\beta}
\Sigma^{\rm (HF)}_{\alpha\beta} \ew{c_\alpha^\dagger c_\beta}' \: .
\ee
Using Wick's theorem ``inversely'',
\be
\Omega_{\ff t,\ff U}^{\rm (HF)}
= 
\Omega_{\ff t',0}
-
\frac{1}{2}
\sum_{\alpha\beta\gamma\delta}
U_{\alpha\beta\delta\gamma} 
\ew{c_{\alpha}^\dagger
c_{\beta}^\dagger
c_\gamma
c_\delta 
}' \: .
\labeq{gphf}
\ee
This is an interesting result as it shows that the Hartree-Fock grand potential is different from the grand potential of the reference system $\Omega_{\ff t',0}$ which is the grand potential of a system of non-interacting electrons.
Due to the ``renormalization'' of the one-particle parameters $\ff t \to \ff t'$, the grand potential of the reference system $\Omega_{\ff t',0}$ does already include some interaction effects. 
As \refeq{gphf} shows, however, there is a certain amount of ``double counting'' of interactions in $\Omega_{\ff t',0}$ which has to be corrected for by the second term.
The second term is the Coulomb interaction energy of the electrons in the renormalized one-particle potential $\ff t'$ and lowers the Hartree-Fock grand potential. 
This is important as we know that $\Omega_{\ff t,\ff U}^{\rm (HF)}$ must represent an upper bound to the exact grand potential of the system:
\be
\Omega_{\ff t',0}
-
\frac{1}{2}
\sum_{\alpha\beta\gamma\delta}
U_{\alpha\beta\delta\gamma} 
\ew{c_{\alpha}^\dagger
c_{\beta}^\dagger
c_\gamma
c_\delta 
}'
\ge
\Omega_{\ff t,\ff U} \: .
\ee

Concluding, we can state that Hartree-Fock theory can very easily be derived from the generalized Ritz principle.
The only approximation consists in the choice of the reference system which serves to span a set of trial density operators. 
The rest of the calculation is straightforward and provides consistent results.

To estimate the quality of the approximation, we replace $U_{\alpha\beta\delta\gamma} \to \lambda U_{\alpha\beta\delta\gamma}$ and expand the {\em exact} grand potential in powers of the interaction strength $\lambda$:
\be
\Omega_{\ff t, \ff U} = \Omega_{\ff t,0}
+
\lambda
\frac{\partial \Omega_{\ff t, 0}}{\partial \lambda} 
+
{\cal O}(\lambda^2) \: .
\ee
Using \refeq{deri}, this gives
\be
\Omega_{\ff t, \ff U} 
= 
\Omega_{\ff t,0}
+
\lambda
\frac{1}{2}
\sum_{\alpha\beta\gamma\delta}
U_{\alpha\beta\delta\gamma} 
\ew{c_{\alpha}^\dagger
c_{\beta}^\dagger
c_\gamma
c_\delta 
}'
+
{\cal O}(\lambda^2) \: ,
\ee
where we have replaced the expectation value with respect to the non-interacting system $H(\ff t,0)$ by the one with respect to the Hartree-Fock reference system $H(\ff t',0)$ where $\ff t'$ is the self-consistent one-particle potential. 
Since 
$
t_{\alpha\beta}
-  
t'_{\alpha\beta}
= -
\sum_{\gamma\delta}
\left(
U_{\alpha\gamma\beta\delta} 
-
U_{\gamma\alpha\beta\delta} 
\right) 
\ew{
c_{\gamma}^\dagger
c_\delta
}'
={\cal O}(\lambda)
$
this is correct up to terms of order $\ca O(\lambda^2)$.
With the same argument we can treat the first term on the r.h.s. yielding:
\be
\Omega_{\ff t, \ff U} 
= 
\Omega_{\ff t',0}
+
\sum_{\alpha\beta} (t_{\alpha\beta} - t'_{\alpha\beta}) 
\frac{\partial \Omega_{\ff t',0}}{\partial t_{\alpha\beta}}
+
{\cal O}(\lambda^2)
+
\lambda
\frac{1}{2}
\sum_{\alpha\beta\gamma\delta}
U_{\alpha\beta\delta\gamma} 
\ew{c_{\alpha}^\dagger
c_{\beta}^\dagger
c_\gamma
c_\delta 
}'
+
{\cal O}(\lambda^2) \: .
\ee
Using \refeq{deri} once more, 
$
\partial \Omega_{\ff t',0} / \partial t_{\alpha\beta} 
= 
\ew{c_\alpha^\dagger c_\beta}'
$, it follows
\be
\Omega_{\ff t, \ff U} 
= 
\Omega_{\ff t',0}
+
\sum_{\alpha\beta} (t_{\alpha\beta} - t'_{\alpha\beta}) 
\ew{
c_{\alpha}^\dagger
c_\beta
}'
+
\frac{1}{2}
\sum_{\alpha\beta\gamma\delta}
U_{\alpha\beta\delta\gamma} 
\ew{c_{\alpha}^\dagger
c_{\beta}^\dagger
c_\gamma
c_\delta 
}'
+
{\cal O}(\lambda^2) \: .
\ee
where we have set $\lambda=1$ in the end.
Comparing with \refeq{hff}, this shows that
\be
\Omega_{\ff t, \ff U} 
= 
\Omega_{\ff t',0}
-
\frac{1}{2}
\sum_{\alpha\beta\gamma\delta}
U_{\alpha\beta\delta\gamma} 
\ew{c_{\alpha}^\dagger
c_{\beta}^\dagger
c_\gamma
c_\delta 
}'
+
{\cal O}(\lambda^2)
\ee
or
\be
\Omega_{\ff t, \ff U} 
= 
\Omega_{\ff t,\ff U}^{\rm (HF)}
+
{\cal O}(\lambda^2)
\: .
\ee
Static mean-field theory thus predicts the correct grand potential of the interacting electron system up to first order in the interaction strength. 
It is easy to see, however, that already at the second order there are deviations. 
In fact, the diagrammatic perturbation theory shows that static mean-field (Hartree-Fock) theory is fully equivalent with self-consistent first-order perturbation theory only.
This leads us to the conclusion that despite its conceptual beauty, static mean-field theory must be expected to give poor results if applied to a system of strongly correlated electrons.

\subsection{Approximation schemes}
\label{sec:types}

\begin{figure}[t]
  \centerline{\includegraphics[width=0.35\textwidth]{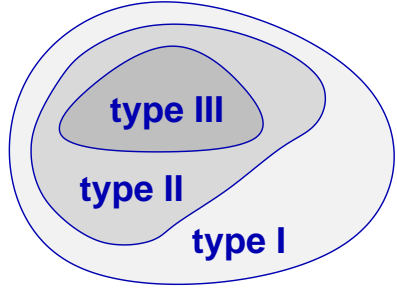}}
\caption{
Hierarchical relation between different types of approximations that can be employed to make use of a formally exact variational principle of the form $\delta \Omega_{\ff t,\ff U}[\ff x] = 0$.
Type I: approximate the Euler equation.
Type II: simplify the functional form.
Type III: restrict the domain of the functional. 
}
\label{fig:types}
\end{figure}

Let us try to learn from the presented construction of static mean-field theory using the generalized Ritz principle and pinpoint the main concepts such that these can be transferred to another variational principle which might then lead to more reliable approximations.
We consider a variational principle of the form
\be
  \delta \widehat{\Omega}_{\ff t,\ff U}[\ff x] = 0 \: ,
\ee
where $\ff x$ is some unspecified multicomponent physical quantity. 
It is assumed that the functional $\widehat{\Omega}_{\ff t,\ff U}[\ff x]$ is stationary at the physical value for $\ff x=\ff x_{\ff t, \ff U}$ and, if evaluated at the physical value, yields the physical grand potential $\Omega_{\ff t, \ff U}$. 
Typically and as we have seen for the Ritz principle, it is generally impossible to exactly evaluate the functional for a given $\ff x$ and that one has to resort to approximations.
Three different types of approximation strategies may be distinguished, see also Fig.\ \ref{fig:types}:

In a {\em type-I approximation} one derives the Euler equation
$\delta \widehat{\Omega}_{\ff t, \ff U}[\ff x] / \delta \ff x = 0$ first and then chooses (a physically motivated) simplification of the equation afterwards to render the determination of $\ff x_{\ff t, \ff U}$ possible.
This is most general but also questionable a priori, as normally the approximated Euler equation no longer derives from some approximate functional.
This may result in thermodynamical inconsistencies.

A {\em type-II approximation} modifies the form of the functional dependence,
$\widehat{\Omega}_{\ff t, \ff U}[\cdots] \to \widehat{\Omega}^{(1)}_{\ff t, \ff U}[\cdots]$, to get a simpler one that allows for a solution of the resulting Euler equation $\delta \widehat{\Omega}^{(1)}_{\ff t, \ff U}[\ff x] / \delta \ff x = 0$.
This type is more particular and yields a thermodynamical potential consistent with $\ff x_{\ff t, \ff U}$.
Generally, however, it is not easy to find a sensible approximation of a functional form.

%*********************************************************************************
%*********************************************************************************

Finally, in a {\em type-III approximation} one restricts the domain of the functional which must then be defined precisely.
This type is most specific and, from a conceptual point of view, should be preferred as compared to type-I or type-II approximations as the exact functional form is retained.
In addition to conceptual clarity and thermodynamical consistency, type-III approximations are truly systematic since improvements can be obtained by an according extension of the domain.

Examples for the different cases can be found e.g.\ in Ref.\ \cite{Pot05}.
The presented derivation of the Hartree-Fock approximation shows that this type-III.
The classification of approximation schemes is hierarchical:
Any type-III approximation can also be understood as a type-II one, and any type-II approximations as type-I, but not vice versa (see Fig.\ \ref{fig:types}).
This does not mean, however, that type-III approximations are superior as compared to type-II and type-I ones.
They are conceptually more appealing but do not necessarily provide ``better'' results.

\section{Dynamical quantities}
\label{sec:5}

To set up the dynamical variational principle, some preparations are necessary. 
We first introduce the one-particle Green's function and the self-energy, briefly sketch diagrammatic perturbation theory and also discuss how within the framework of perturbation theory it is possible to construct non-perturbative approximations like the so-called cluster perturbation theory (CPT).

\subsection{Green's functions}

We again consider a system of interacting electrons in thermodynamical equilibrium at temperature $T$ and chemical potential $\mu$. 
The Hamiltonian of the system is $H = H(\ff t, \ff U)$, see \refeq{mod}.
Now, the one-particle Green's function \cite{AGD64,FW71,NO88}
\be
G_{\alpha\beta}(\omega) \equiv \langle\langle c_\alpha ; c_\beta^\dagger \rangle \rangle
\ee
of the system will be the main object of interest. 
This is a frequency-dependent quantity which provides information on the static expectation value of the one-particle density matrix $c^\dagger_\alpha c_\beta$ but also on the spectrum of one-particle excitations related to a photoemission experiment \cite{Pot01b}.
The Green's function can be defined for complex $\omega$ via its spectral representation:
\be
  G_{\alpha\beta}(\omega) = \int_{-\infty}^\infty dz \: \frac{A_{\alpha\beta}(z)}{\omega -  z} \: ,
\labeq{spectralrep}
\ee
where the spectral density 
\be
A_{\alpha\beta}(z) = \int_{-\infty}^\infty dt \: \exp(izt) A_{\alpha\beta}(t)
\ee
is the Fourier transform of 
\be
  A_{\alpha\beta}(t) = \frac{1}{2\pi} \langle [c_\alpha(t) , c^\dagger_\beta(0)]_+ \rangle \; ,
\ee
which involves the anticommutator of an annihilator and a creator with a Heisenberg time dependence $O(t) = \exp(i(H-\mu N)t) O \exp(- i(H-\mu N)t)$.
Due to the thermal average, $\langle O \rangle =  \tr (\rho_{\ff t, \ff U} O) / Z_{\ff t,\ff U}$, the Green's function parametrically depends on $\ff t$ and $\ff U$ and is denoted by $G_{\ff t, \ff U ,\alpha\beta}(\omega)$.

For the diagram technique employed below, we need the Green's function on the imaginary Matsubara frequencies $\omega = i \omega_n \equiv i(2n+1)\pi T$ with integer $n$ \cite{AGD64,FW71,NO88}.
In the following the elements $G_{\ff t, \ff U ,\alpha\beta}(i\omega_n)$ are considered to form a matrix $\ff G_{\ff t, \ff U}$ which is diagonal with respect to $n$.

The ``free'' Green's function $\ff G_{\ff t, 0}$ is obtained for $\ff U = 0$, and its elements are given by:
\be
  G_{\ff t, 0, \alpha\beta}(i\omega_n) = 
  \left(\frac{1}{i\omega_n + \mu - \ff t}\right)_{\alpha\beta} \: .
\ee
This is a result that can easily be derived from the equation-of-motion technique \cite{FW71}.
Therewith, we can define the self-energy via Dyson's equation
\be
  \ff G_{\ff t, \ff U} = \frac{\ff 1}{\ff G_{\ff t, 0}^{-1} - \ff \Sigma_{\ff t, \ff U}} \; , 
\labeq{dyson}
\ee
i.e.\ $\ff \Sigma_{\ff t, \ff U} = \ff G_{\ff t, 0}^{-1} - \ff G_{\ff t, \ff U}^{-1}$.
The full meaning of this definition becomes clear within the context of diagrammatic perturbation theory \cite{AGD64,FW71,NO88}.

\subsection{Diagrammatic perturbation theory}

The main reason why the Green's function is put into the focus of the theory is that a systematic expansion in powers of the interaction can be set up.
Here, a brief sketch of perturbation theory can be given only (see Refs.\ \cite{AGD64,FW71,NO88} for details).
Starting point is the so-called S-matrix defined for $0\le \tau, \tau' \le \beta$ as 
\be
  S(\tau, \tau') = e^{\ca H_0 \tau} e^{- \ca H (\tau - \tau')} e^{-\ca H_0\tau'} \; ,
\labeq{sm}
\ee
where $\ca H=H_{\ff t,\ff U} -\mu N$ and $\ca H_0=H_{\ff t,0} - \mu N$. 

There are two main purposes of the S-matrix.
First, it serves to rewrite the partition function in the following way:
\be
Z _{\ff t,\ff U}
= 
\tr e^{-\beta \ca H}
=
\tr \left(
 e^{-\beta \ca H_0 }
 e^{\beta \ca H_0}
 e^{-\beta \ca H}
\right)
= 
\tr \left(
 e^{-\beta \ca H_0}
 S(\beta,0)
\right)
= 
 Z_{\ff t,0}
 \ew{
 S(\beta,0)
 }^{(0)} \: .
\ee
The partition function of the interacting system is thereby given in terms of the partition function of the {\em free} system, which is known, and a {\em free} expectation value of the S-matrix.

The second main purpose is related to the time dependence. 
The Green's function on the Matsubara frequencies $G_{\alpha\beta}(i\omega_n)$ is related to the imaginary-time Green's function $G_{\alpha\beta}(\tau)$ via discrete Fourier transformation: 
\be
G_{\alpha\beta}(\tau) = \frac{1}{\beta} \sum_{n=-\infty}^\infty G_{\alpha\beta}(i\omega_n)  
\, e^{-i \omega_n \tau}
\ee
and 
\be
G_{\alpha\beta}(i\omega_n) = \int_0^\beta d\tau \, G_{\alpha\beta}(\tau) \, e^{i\omega_n \tau} \: .
\ee
Let $\ca T$ be the (imaginary) time-ordering operator.
Then
\be
G_{\alpha\beta}(\tau) 
=
-
\langle
\ca T
c_\alpha(\tau) c_\beta^\dagger(0)
\rangle
\ee
is given in terms of an annihilator or creator which has the full (imaginary) Heisenberg time dependence:
\be
  c_\alpha(\tau) = e^{\ca H \tau} c_\alpha e^{-\ca H \tau} \: .
\ee
With the help of the S-matrix the interacting time dependence can be transformed into a {\em free} time dependence, namely:
\be
c_\alpha(\tau)  
= 
S(0,\tau) c_{I,\alpha}(\tau) S(\tau,0)
\; , \qquad
c^\dagger_{\alpha}(\tau)  
= 
S(0,\tau) c^\dagger_{I,\alpha}(\tau) S(\tau,0) \: .
\ee
Here, the index $I$ (``interaction picture'') indicates that the time dependence is due to $\ca H_0$ only. 
This time dependence is simple and known:
\be
c_{I,\alpha}(\tau)
=
\sum_{\beta}
\left(
e^{-(\ff t -\mu)\tau}
\right)_{\alpha\beta}
c_\beta
\; , \qquad
c^\dagger_{I,\alpha}(\tau)
=
\sum_{\beta}
\left(
e^{+(\ff t -\mu)\tau}
\right)_{\alpha\beta}
c^\dagger_\beta \: .
\ee

Therefore, the remaining problem consists in the determination of the S-matrix itself. 
The defining equation \refeq{sm} cannot be used since this involves the exponential of the fully interacting Hamiltonian. 
There is, however, a much more suitable representation of the S-matrix. 
To derive this, we first set up its equation of motion. 
A straightforward calculation shows:
\be
- \frac{\partial}{\partial \tau} S(\tau,\tau') = H_{1,I}(\tau) S(\tau,\tau') \: .
\labeq{dgl}
\ee
Here, $H_1 = H - H_0$ is the interaction part of the Hamiltonian and in $H_{1,I}(\tau)$ the time dependence is due to $H_0$ only. 
A formal solution of this differential equation with the initial condition $S(\tau, \tau) = 1$ can be derived easily using the time-ordering operator $\ca T$ again:
\be
  S(\tau,\tau') = \ca T \exp \left( - \int_{\tau'}^\tau \, d\tau'' V_I(\tau'') \right) \: .
\labeq{sr}
\ee
Note, that if all quantities were commuting, the solution of \refeq{dgl} would trivially be given by \refeq{sr} without $\ca T$. 
The appearance of $\ca T$ can therefore be understood as necessary to enforce commutativity.

Using this S-matrix representation, the partition function and the Green's function can be written as:
\be
\frac{Z}{Z_0}
=
 \Big\langle
 \ca T \exp \left( - \int_{0}^\beta \, d\tau'' V_I(\tau'') \right)
 \Big\rangle^{(0)}
\labeq{p1}
\ee
and
\be
G_{\alpha\beta}(\tau)
=
- 
\frac{ \Big\langle 
\ca T \exp \left( - \int_{0}^\beta \, d\tau V(\tau) \right)
c_{\alpha}(\tau) c_{\beta}^\dagger(0) \Big\rangle^{(0)} }
{
\Big\langle
 \ca T \exp \left( - \int_{0}^\beta \, d\tau V(\tau) \right)
 \Big\rangle^{(0)}
} \: .
\labeq{p2}
\ee
The important point is that the expectation values and time dependencies appearing here are free and thus known. 

Expanding the respective exponentials in \refeq{p1} and \refeq{p2} provides an expansion of the static partition function and of the dynamic Green's function in powers of the interaction strength. 
The coefficients of this expansion are given as free expectation values of time-ordered products of annihilators and creators with free time dependence. 
Hence, according to Wick's theorem the coefficient at the $k$-th order is given by a sum of $(2k)!$ terms (in case of the partition function, for example), each of which factorizes into an $k$-fold product of terms of the form $\langle \ca T c_\alpha(\tau) c^\dagger_{\alpha'}(\tau') \rangle^{(0)}$ called contractions.
Apart from a sign, this contraction is nothing but the free Green's function. 
The summation of the $(2k)!$ terms is organized by means of a diagrammatic technique where vertices ($\ff U$) are linked via propagators ($\ff G_{\ff t, 0}$).
The details of this technique can be found in Refs.\ \cite{AGD64,FW71,NO88}, for example.

%*******************************************************************************
\begin{figure}[t]
\includegraphics[width=85mm]{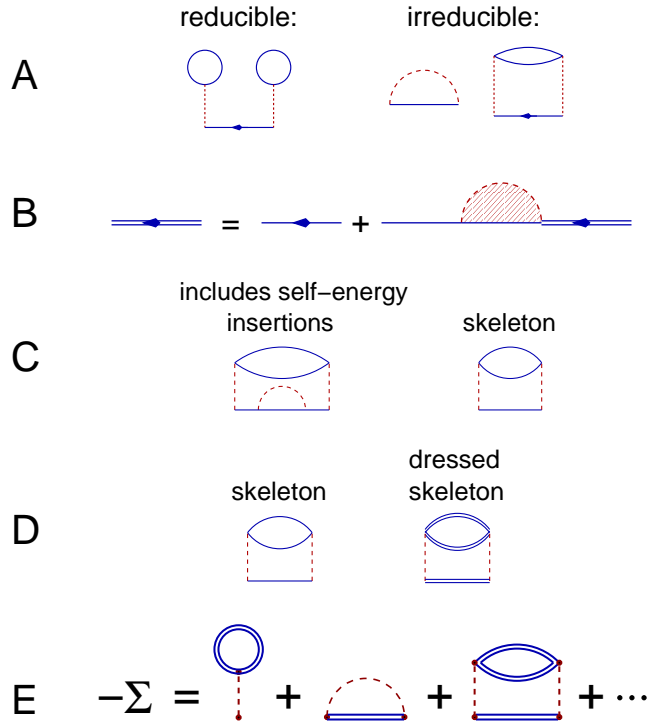}
\centering
\caption{
(A) A reducible and two irreducible self-energy insertions. 
Vertices ($\ff U$) are represented by dashed lines, propagators ($\ff G_{\ff t,0}$) by solid lines.
(B) Diagrammatic representation of Dyson's equation. 
The fully interacting Green's function is symbolized by a double line, the self-energy by a semi-circle.
(C) Self-energy diagrams with (left) and without (right) self-energy insertions. 
(D) A skeleton self-energy diagram (left) and a dressed skeleton (right).
(E) The skeleton-diagram expansion of the self-energy.
}
\label{fig:sigma}
\end{figure}
%*******************************************************************************

There are three important points, however, which are worthwhile to be mentioned here:
The first is the linked-cluster theorem which allows us to concentrate on connected diagrams only.
In the case of the Green's function, the disconnected diagrams exactly cancel the denominator in \refeq{p2}. 
As concerns closed diagrams contributing to the partition function, \refeq{p1}, the sum of only the connected closed diagrams yields, apart from a constant, $\ln Z_{\ff t,\ff U}$, i.e.\ the grand potential.

Second, we can identify so-called self-energy insertions in the diagrammatic series, i.e.\ parts of diagrams with exhibit links to two external propagators. 
Examples are given in Fig.\ \ref{fig:sigma}A where we also distinguish between reducible and irreducible self-energy insertions. 
The reducible ones can be split into two disconnected parts by removal of a single propagator line.
The self-energy is then defined diagrammatically as the sum over all irreducible self-energy insertions, and Dyson's equation can be derived diagrammatically, see Fig.\ \ref{fig:sigma}B.

Third, there are obviously irreducible self-energy diagrams which contain self-energy insertions, see the first diagram in Fig.\ \ref{fig:sigma}C, for example.
Self-energy diagrams without any self-energy insertion are called skeleton diagrams. 
Skeleton diagrams can be ``dressed'' by replacing in the diagram the propagators which stand for the free Green's function with ``full'' propagators (double lines) which stand for the fully interacting Green's function, see Fig.\ \ref{fig:sigma}D.
It is easy to see that the self-energy is given by the sum of the skeleton diagrams only provided that these are dressed, see Fig.\ \ref{fig:sigma}E.
Therewith, the self-energy is given in terms of the interacting Green's function. 
The corresponding functional relationship is called skeleton-diagram expansion. 

Some more important properties of the self-energy can be listed:
(i) The self-energy has a spectral representation similar to \refeq{spectralrep}.
(ii) In particular, the corresponding spectral function (matrix) is positive definite, and the poles of
$\ff \Sigma_{\ff t, \ff U}$ are on the real axis \cite{Lut61}.
(iii) $\Sigma_{\alpha\beta}(\omega) = 0$ if $\alpha$ or $\beta$ refer to one-particle orbitals that are non-interacting, i.e.\ if $\alpha$ or $\beta$ do not occur as an entry of the matrix of interaction parameters $\ff U$. 
Those orbitals or sites are called non-interacting. 
This property of the self-energy is clear from its diagrammatic representation.
(iv) If $\alpha$ refers to the sites of a Hubbard-type model with local interaction, the self-energy can generally be assumed to be more local than the Green's function. 
This is corroborated e.g.\ by explicit calculations using weak-coupling perturbation theory \cite{SC90,SC91,PN97c} and by the fact that the self-energy is purely local on infinite-dimensional lattices \cite{MV89,MH89b}.

\subsection{Cluster perturbation theory}

The method of Green's functions and diagrammatic perturbation theory represents a powerful approach to study systems of interacting electrons in thermal equilibrium. 
At first sight it seems that only weakly interacting systems are accessible with this technique.
In fact, summing only the diagrams up to a given order in the expansion represents a strict weak-coupling approach.
For prototypical lattice models with local interaction, such as the Hubbard model, however, an expansion in powers of the hopping appears to be more attractive since most of the interesting phenomena, like collective magnetism or correlation-driven metal-insulator transitions emerge in the strong-coupling regime.

%*******************************************************************************
\begin{figure}[t]
\includegraphics[width=85mm]{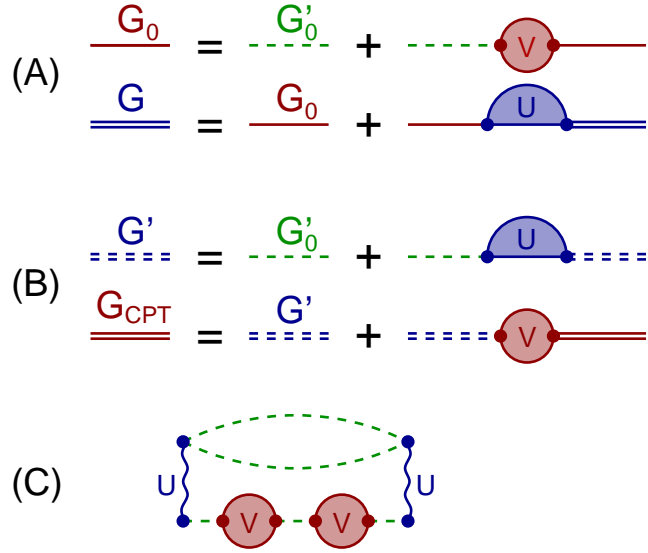}
\centering
\caption{
To derive the CPT, consider perturbation theory in $U$ and the inter-cluster hopping $V$. 
(A) The exact Green's function would be obtained by starting from the free ($U=V=0$) propagator $\ff G'_0$ and, in a first step, sum the diagrams of all orders in $V$ but for $U=0$.
Since the diagrammatic series is a geometrical series only, this can be done exactly. 
Formally, this corresponds to usual scattering theory for free electrons and yields the $U=0$ Green's function $\ff G_0$.
In a second step, the diagrams generated by electron-electron interaction $U$ would have to be summed up to get the full $\ff G$.
This, however, cannot be done in practice. 
(B) The CPT approximation reverses the two above steps.
First, the free ($U=V=0$) propagator $\ff G'_0$ is renormalized by electron-electron interaction $U$ to all orders but at $V=0$. 
This yields the fully interacting cluster Green's function $\ff G'$. 
While, of course, $\ff G'$ cannot be computed by the extremely complicated summation of individual $U$ diagrams, it is easily accessible numerically if the cluster size is sufficiently small.
In the second step, the $V=0$ propagator $\ff G'$ is renormalized due to potential scattering.
This yields an approximation $\ff G_{\rm CPT}$ to the exact Green's function $\ff G$. 
(C) The CPT treatment is approximate since certain diagrams are neglected. 
The last panel shows a self-energy diagram, second order in $U$, second order in $V$, which is not taken into account within CPT.
}
\label{fig:cpt}
\end{figure}
%*******************************************************************************

Let us therefore focus on the Hubbard model in the atomic limit for a moment:
\be
  H_{\rm at} = \sum_{i\sigma} t_{ii} c^\dagger_{i\sigma}c_{i\sigma} + \frac{U}{2} \sum_{i\sigma} n_{i\sigma} n_{i-\sigma} \: .
\ee
Here, the problem separates into atomic problems which can be solved easily since the atomic Hilbert space is small.
It is therefore tempting to start from the atomic limit and to treat the rest of the problem, the ``embedding'' of the atom into the lattice, in some approximate way. 
The main idea of the so-called Hubbard-I approximation \cite{Hub63} is to calculate the one-electron Green's function from the Dyson equation where the self-energy is approximated by the self-energy of the atomic system.
This is one of the most simple embedding procedures. 
It already shows that the language of diagrammatic perturbation theory, Green's functions and diagrammatic objects, such as the self-energy, can be very helpful to construct an embedding scheme.

The Hubbard-I approach turns out to be a too crude approximation to describe the above-mentioned collective phenomena. 
One of its advantages, however, is that it offers a perspective for systematic improvement:
Nothing prevents us to start with a more complicated ``atom'' and employ the same trick:
We consider a partition of the underlying lattice with $L$ sites (where $L \to \infty$) into $L/L_c$ disconnected clusters consisting of $L_c$ sites each. 
If $L_c$ is not too large, the self-energy of a single Hubbard cluster is accessible by standard numerical means \cite{Dag94} and can be used as an approximation in the Dyson equation to get the Green's function of the full model. 
This leads to the cluster perturbation theory (CPT) \cite{GV93,SPPL00}.  

CPT can also be motivated by treating the Hubbard interaction $U$ and the {\em inter}-cluster hopping $V$ as a perturbation of the system of disconnected clusters with {\em intra}-cluster hopping $t'$.
This has been shown recently \cite{BP11}.
The CPT Green's function is then obtained by summing the diagrams in perturbation theory to all orders in $U$ and $V$ but neglecting vertex corrections which intermix $U$ and $V$ interactions, see Fig.\ \ref{fig:cpt}.

While these two ways of deriving CPT are equivalent, one aspect of the former is interesting: 
Taking the self-energy from some reference model (the cluster) is reminiscent of dynamical mean-field theory (DMFT) \cite{MV89,GKKR96,KV04} where the self-energy of an impurity model approximates the self-energy of the lattice model. 
This provokes the question whether both, the CPT and the DMFT, can be understood in single unifying theoretical framework.

This question is another motivation for the construction of approximations from a dynamical variational principle. Yet another one is that there are certain deficiencies of the CPT.
While CPT can be seen as a cluster mean-field approach since correlations beyond the cluster extensions are neglected, it not self-consistent, i.e.\ there is no feedback of the resulting Green's function on the cluster to be embedded (some {\em ad hoc} element of self-consistency is included in the original Hubbard-I approximation).
In particular, there is no concept of a Weiss mean field as it is the case within static mean-field theory, and, therefore, CPT can neither describe different phases of an extended system nor phase transitions.
Another related point is that CPT provides the Green's function only but no thermodynamical potential.
Different ways to derive e.g.\ the free energy from the Green's function \cite{AGD64,FW71,NO88} give inconsistent results.

\section{Self-energy-functional theory}
\label{sec:6}

To overcome these deficiencies, a self-consistent cluster-embedding scheme has to be set up. 
Ideally, this results from a variational principle for a general thermodynamical potential which is formulated in terms of dynamical quantities as e.g.\ the self-energy or the Green's function.
The variational formulation should ensure the internal consistency of corresponding approximations and should make contact with the DMFT. 
These points represent the goals self-energy-functional theory \cite{Pot03a,Pot03b,PAD03,Pot05} which will be developed in the following.

\subsection{Luttinger-Ward generating functional}

First of all, we would like to distinguish between dynamic quantities, like the self-energy, which is frequency-dependent and related to the (one-particle) excitation spectrum, on the one hand, and static quantities, like the grand potential and its derivatives with respect to $\mu$, $T$, etc.\ which are related to the thermodynamics, on the other. 
A link between static and dynamic quantities is needed to set up a variational principle which gives the (dynamic) self-energy by requiring a (static) thermodynamical potential be stationary, 
There are several such relations \cite{AGD64,FW71,NO88}.
The Luttinger-Ward functional $\widehat{\Phi}_{\ff U}[\ff G]$ provides a special relation with several advantageous properties \cite{LW60}: 
\bi
\item[](i) 
$\widehat{\Phi}_{\ff U}[\ff G]$ is a functional.
Functionals $\widehat{A} = \widehat{A}[\cdots]$ are indicated by a hat and should be distinguished clearly from physical quantities $A$.
\\

\item[](ii)
The domain of the Luttinger-Ward functional is given by ``the space of Green's functions''. 
This has to be made more precise later. 
\\

\item[](iii)
If evaluated at the exact (physical) Green's function, $\ff G_{\ff t, \ff U}$, of the system with Hamiltonian $H = H(\ff t, \ff U)$, the Luttinger-Ward functional gives a quantity
\be
  \widehat{\Phi}_{\ff U}[\ff G_{\ff t, \ff U}] = \Phi_{\ff t, \ff U}
\labeq{phiphys}
\ee
which is related to the grand potential of the system via:
\be
  \Omega_{\ff t, \ff U}  
  = 
  \Phi_{\ff t, \ff U} 
  +
  \Tr \ln \ff G_{\ff t, \ff U} 
  - 
  \Tr ( \ff \Sigma_{\ff t, \ff U} \ff G_{\ff t, \ff U}) \; .
\labeq{phiom}
\ee
Here the notation 
\be
\Tr \, \ff A \equiv T \sum_n \sum_\alpha e^{i\omega_n0^+} A_{\alpha\alpha}(i\omega_n)
\ee
is used where $0^+$ is a positive infinitesimal.
\\

\item[](iv)
The functional derivative of the Luttinger-Ward functional with respect to its argument is:
\be
  \frac{1}{T} \frac{\delta \widehat{\Phi}_{\ff U}[\ff G]}{\delta \ff G} 
  = \widehat{\ff \Sigma}_{\ff U}[\ff G] \: .
\label{eq:der}
\ee
Clearly, the result of this operation is a functional of the Green's function again. 
This functional is denoted by $\widehat{\ff \Sigma}$ since its evaluation at the physical (exact) Green's function $\ff G_{\ff t, \ff U}$ yields the physical self-energy:
\be
  \widehat{\ff \Sigma}[\ff G_{\ff t, \ff U}] = \ff \Sigma_{\ff t, \ff U}  \: .
\labeq{sigmaskel}
\ee

\item[](v)
The Luttinger-Ward functional is ``universal'': 
The functional relation $\widehat{\Phi}_{\ff U}[\cdots]$ is completely determined by the interaction parameters $\ff U$ (and does not depend on $\ff t$).
This is made explicit by the subscript.
Two systems (at the same chemical potential $\mu$ and temperature $T$) with the same interaction $\ff U$ but different one-particle parameters $\ff t$ (on-site energies and hopping integrals) are described by the same Luttinger-Ward functional. 
Using Eq.\ (\ref{eq:der}), this implies that the functional $\widehat{\ff \Sigma}_{\ff U}[\ff G]$ is universal, too.
\\

\item[](vi)
Finally, the Luttinger-Ward functional vanishes in the non-interacting limit:
\be
  \widehat{\Phi}_{\ff U}[\ff G] \equiv 0  \quad \mbox{for} \quad \ff U = 0 \; .
\ee
\ei

\subsection{Diagrammatic derivation}

\begin{figure}[t]
  \centerline{\includegraphics[width=0.55\textwidth]{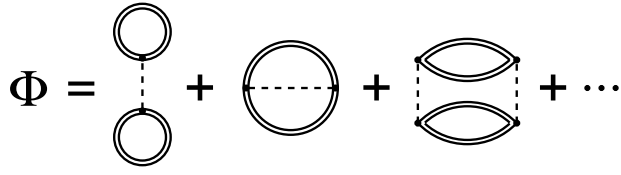}}
\caption{
Original definition of the Luttinger-Ward functional $\widehat{\Phi}_{\ff U}[\ff G]$, see Ref.\ \cite{LW60}.
Double lines: fully interacting propagator $\ff G$. 
Dashed lines: interaction $\ff U$.
}
\label{fig:lw}
\end{figure}

In the original paper by Luttinger and Ward \cite{LW60} it is shown that $\widehat{\Phi}_{\ff U}[\ff G]$ can be constructed order by order in diagrammatic perturbation theory.
The functional is obtained as the limit of the infinite series of closed renormalized skeleton diagrams, i.e.\
closed diagrams without self-energy insertions and with all free propagators replaced by fully interacting ones (see Fig.\ \ref{fig:lw}). 
There is no known case where this skeleton-diagram expansion could be summed up to get a closed form for $\widehat{\Phi}_{\ff U}[\ff G]$.
Therefore, the explicit functional dependence is unknown even for the most simple types of interactions like the Hubbard interaction.

Using the classical diagrammatic definition of the Luttinger-Ward functional, the properties (i) -- (vi) listed in the previous section are easily verified:
By construction, $\widehat{\Phi}_{\ff U}[\ff G]$ is a functional of $\ff G$ which is universal (properties (i), (ii), (v)).
Any diagram is the series depends on $\ff U$ and on $\ff G$ {\em only}.
Particularly, it is independent of $\ff t$.
Since there is no zeroth-order diagram, $\widehat{\Phi}_{\ff U}[\ff G]$ trivially vanishes for 
$\ff U=0$, this proves (vi).

Diagrammatically, the functional derivative of $\widehat{\Phi}_{\ff U}[\ff G]$ with respect to $\ff G$ corresponds to the removal of a propagator from each of the $\Phi$ diagrams. 
Taking care of topological factors \cite{LW60,AGD64}, one ends up with the skeleton-diagram expansion of the self-energy (iv), see also Fig.\ \ref{fig:sigma}E.
Therefore, Eq.\ (\ref{eq:sigmaskel}) is obtained in the limit of this expansion. 

\refeq{phiom} can be derived by a coupling-constant integration as has been shown in Ref.\ \cite{LW60}.
Alternatively, it can be verified by integrating over $\mu$.
To this end, we first consider the $\mu$ derivative of the different terms on the r.h.s.\ of \refeq{phiom}. 
For the first term we have:
\be
\frac{\partial}{\partial \mu} \Phi_{\ff t, \ff U} 
=
\frac{\partial}{\partial \mu} \widehat{\Phi}_{\ff U}[\ff G_{\ff t,\ff U}]
=
\sum_{\alpha\beta}\sum_n
\frac{\delta \widehat{\Phi}_{\ff U}[\ff G_{\ff t, \ff U}]}{\delta G_{\alpha\beta}(i\omega_n)} 
\frac{\partial G_{\ff t,\ff U;\alpha\beta}(i\omega_n)}{\partial \mu}
\: ,
\ee
and thus
\be
\frac{\partial}{\partial \mu} \Phi_{\ff t, \ff U} 
=
\sum_{\alpha\beta}T\sum_n
\Sigma_{\ff t,\ff U;\beta\alpha}(i\omega_n)
\frac{\partial G_{\ff t, \ff U;\alpha\beta}(i\omega_n)}{\partial \mu}
=
\mbox{Tr} \left( \ff \Sigma_{\ff t,\ff U} \frac{\partial \ff G_{\ff t,\ff U}}{\partial \mu} \right)
\: .
\ee
For the second one, we find:
\be
\frac{\partial}{\partial \mu} \mbox{Tr} \ln \ff G_{\ff t,\ff U} 
=
\mbox{Tr} \left( \ff G_{\ff t,\ff U}^{-1} \frac{\partial \ff G_{\ff t,\ff U} }{\partial \mu} \right)
\; ,
\ee
and for the third:
\be
\frac{\partial}{\partial \mu} \mbox{Tr} (\ff \Sigma_{\ff t,\ff U} \ff G_{\ff t,\ff U})
=
\mbox{Tr} \left(\frac{\partial \ff \Sigma_{\ff t,\ff U}}{\partial \mu} \ff G_{\ff t,\ff U} \right)
+
\mbox{Tr} \left(\ff \Sigma_{\ff t,\ff U} \frac{\partial \ff G_{\ff t,\ff U}}{\partial \mu} \right)
\: .
\ee
Next, we note that
\ba
&&\Tr \left( \ff G_{\ff t, \ff U}^{-1} \frac{\partial \ff G_{\ff t, \ff U}}{\partial \mu} \right)
- \Tr \left( \ff G_{\ff t, \ff U} \frac{\partial \ff \Sigma_{\ff t, \ff U}}{ 
\partial \mu} \right)
= 
\mbox{Tr} \left[\left( \ff G_{\ff t, \ff U}^{-1} \frac{\partial \ff G_{\ff t, \ff U}}{\partial \mu} \ff G_{\ff t, \ff U}^{-1} - \frac{\partial \ff \Sigma_{\ff t, \ff U}}{\partial \mu}\right) \ff G_{\ff t, \ff U} \right]
\nonumber \\
&&
= 
\mbox{Tr} \left[ \frac{\partial ( - \ff G_{\ff t, \ff U}^{-1} - \ff \Sigma_{\ff t, \ff U})}{\partial \mu} \ff G_{\ff t, \ff U} \right]
=
- \Tr \: \ff G_{\ff t, \ff U} 
\: ,
\ea
where Dyson's equation has been used in the last step and $\partial \ff G_{\ff t, 0}^{-1} / \partial \mu = \ff 1$.
Collecting the results, we have:
\be
\frac{\partial}{\partial \mu} (\Phi_{\ff t, \ff U} + \Tr \ln \ff G_{\ff t, \ff U} 
- \Tr \, \ff \Sigma_{\ff t, \ff U} \ff G_{\ff t, \ff U}) 
= - \Tr \: \ff G_{\ff t, \ff U} 
\ee
From the definition of the Green's function we immediately have
\be
  \Tr \: \ff G_{\ff t, \ff U} = \langle N \rangle_{\ff t, \ff U} \: .
\ee 
$\langle N \rangle_{\ff t, \ff U}$ is the grand-canonical average of the total particle-number operator.
Apart from a sign, this can be written as the $\mu$ derivative of the grand potential.
Hence,
\be
\frac{\partial}{\partial \mu} (\Phi_{\ff t, \ff U} + \Tr \ln \ff G_{\ff t, \ff U} 
- \Tr \, \ff \Sigma_{\ff t, \ff U} \ff G_{\ff t, \ff U}) 
= \frac{\partial \Omega_{\ff t, \ff U} }{ \partial \mu} \: . 
\ee
Integration over $\mu$ then yields \refeq{phiom}.
Note that the equation holds trivially for $\mu \to -\infty$, i.e.\ for 
$\langle N \rangle_{\ff t, \ff U} \to 0$ since $\ff \Sigma_{\ff t, \ff U} = 0$ 
and $\Phi_{\ff t, \ff U} =0$ in this limit.

%*********************************************************************************
%*********************************************************************************

\subsection{Derivation using the path integral}

For the diagrammatic derivation it has to be assumed that the skeleton-diagram series is convergent. 
It is therefore interesting to see how the Luttinger-Ward functional can be defined and how its properties can be verified within a path-integral formulation.
This is essentially non-perturbative.
The path-integral construction of the Luttinger-Ward functional was first given in Ref.\ \cite{Pot06b}.

Using Grassmann variables \cite{NO88}
$\xi_\alpha(i\omega_n)= T^{1/2} \int_0^{1/T} d\tau \: e^{i\omega_n\tau} \xi_\alpha(\tau)$ and $\xi^\ast_\alpha(i\omega_n)= T^{1/2} \int_0^{1/T} d\tau \: e^{-i\omega_n\tau} \xi^\ast_\alpha(\tau)$, the elements of the Green's function are given by 
$G_{\ff t, \ff U, \alpha\beta}(i\omega_n) = - \langle \xi_\alpha(i\omega_n) \xi^\ast_\beta(i\omega_n) \rangle_{\ff t, \ff U}$. 
The average
\be
  G_{\ff t, \ff U, \alpha\beta}(i\omega_n) 
  =
  \frac{-1}{Z_{\ff t,\ff U}}
  \int D \xi D \xi^\ast 
  \xi_\alpha(i\omega_n) \xi^\ast_\beta(i\omega_n)
  \exp\left( A_{\ff t, \ff U,\xi\xi^\ast} \right)
\labeq{gfunct}
\ee
is defined with the help of the action
$A_{\ff t, \ff U,\xi\xi^\ast} = A_{\ff t,\xi\xi^\ast} + A_{\ff U,\xi\xi^\ast}$
where
\be
   A_{\ff t, \ff U,\xi\xi^\ast} 
   =
   \sum_{n,\alpha\beta} \xi_\alpha^\ast(i\omega_n) 
   ((i\omega_n + \mu)\delta_{\alpha\beta} - t_{\alpha\beta})
   \xi_\beta(i\omega_n) 
   +
   A_{\ff U,\xi\xi^\ast}
\ee
and 
\be
   A_{\ff U,\xi\xi^\ast}
   = 
   - \frac{1}{2} \sum_{\alpha\beta\gamma\delta} U_{\alpha\beta\delta\gamma} \int_0^{1/T} 
   \!\! d\tau \:
   \xi_\alpha^\ast(\tau) 
   \xi_\beta^\ast(\tau)
   \xi_\gamma(\tau) 
   \xi_\delta(\tau) \; .
\ee
This is the standard path-integral representation of the Green's function \cite{NO88}.

The action can be considered as the physical action which is obtained when evaluating the {\em functional}
\be
   \widehat{A}_{\ff U,\xi\xi^\ast}[\ff G_0^{-1}] =
   \sum_{n,\alpha\beta} \xi_\alpha^\ast(i\omega_n) 
   G_{0,\alpha\beta}^{-1}(i\omega_n)
   \xi_\beta(i\omega_n) 
   +
   A_{\ff U,\xi\xi^\ast} \: 
\ee
at the (matrix inverse of the) physical free Green's function, i.e.\
\be
   A_{\ff t, \ff U,\xi\xi^\ast} 
   = 
   \widehat{A}_{\ff U,\xi\xi^\ast}[\ff G_{\ff t,0}^{-1}] \; .
\ee
Using this, we define the functional
\be
  \widehat{\Omega}_{\ff U}[\ff G_0^{-1}] =  - T \ln \widehat{Z}_{\ff U}[\ff G_0^{-1}]
\labeq{omgfree}
\ee
with
\be
  \widehat{Z}_{\ff U}[\ff G_0^{-1}] = \int D \xi D \xi^\ast 
  \exp\left( \widehat{A}_{\ff U,\xi\xi^\ast}[\ff G_0^{-1}] \right) \; .
\ee
The functional dependence of $\widehat{\Omega}_{\ff U}[\ff G_0^{-1}]$ is determined by $\ff U$ only, i.e.\ the functional is universal. 
Obviously, the physical grand potential is obtained when inserting the physical inverse free Green's function
$\ff G_{\ff t, 0}^{-1}$:
\be
  \widehat{\Omega}_{\ff U}[\ff G_{\ff t, 0}^{-1}] = \Omega_{\ff t, \ff U} \: .
\labeq{omex0} 
\ee
The functional derivative of \refeq{omgfree} leads to another universal functional:
\be
\frac{1}{T} \frac{\delta \widehat{\Omega}_{\ff U}[\ff G_0^{-1}]}
{\delta \ff G_0^{-1}} 
= - \: \frac{1}{\widehat{Z}_{\ff U}[\ff G_0^{-1}]} 
\frac{\delta \widehat{Z}_{\ff U}[\ff G_0^{-1}]}
{\delta \ff G_0^{-1}} 
\equiv - \widehat{\cal \ff G}_{\ff U}[\ff G_0^{-1}] \: ,
\labeq{omder}
\ee
with the property
\be
  \widehat{\cal \ff G}_{\ff U}[\ff G_{\ff t, 0}^{-1}] = \ff G_{\ff t, \ff U} \: .
\labeq{calg}
\ee
This is easily seen from \refeq{gfunct}.

The strategy to be pursued is the following: 
$\widehat{\cal \ff G}_{\ff U}[\ff G_{0}^{-1}]$ is a universal ($\ff t$ independent) functional and can be used to construct a universal relation $\ff \Sigma = \widehat{\ff \Sigma}_{\ff U}[\ff G]$ 
between the self-energy and the one-particle Green's function -- independent from the Dyson equation (\ref{eq:dyson}).
Using this and the universal functional $\widehat{\Omega}_{\ff U}[\ff G_{0}^{-1}]$, a universal functional $\widehat{\Phi}_{\ff U}[\ff G]$ can be constructed, the derivative of which essentially yields $\widehat{\ff \Sigma}_{\ff U}[\ff G]$ and that also obeys all other properties of the diagrammatically constructed Luttinger-Ward functional.
This also means that $\widehat{\ff \Sigma}_{\ff U}[ \ff G ]$ corresponds to the skeleton-diagram expansion of the self-energy.

To start with, consider the equation
\be
  \widehat{\cal \ff G}_{\ff U}[ \ff G^{-1} + \ff \Sigma ] = \ff G \: .
\labeq{rel}  
\ee
This is a relation between the variables $\ff \Sigma$ and $\ff G$ which
for a given $\ff G$, may be solved for $\ff \Sigma$. 
This defines a functional $\widehat{\ff \Sigma}_{\ff U}[\ff G]$, i.e.
\be
  \widehat{\cal \ff G}_{\ff U}[ \ff G^{-1} - \widehat{\ff \Sigma}_{\ff U}[\ff G] ] = \ff G \: .
\labeq{gfunc}
\ee
For a given Green's function $\ff G$, the self-energy $\ff \Sigma = \widehat{\ff \Sigma}_{\ff U}[\ff G]$ is defined to be the solution of \refeq{rel}.
From the Dyson equation (\ref{eq:dyson}) and \refeq{calg} it is obvious that the relation (\ref{eq:rel}) is satisfied for the physical $\ff \Sigma = \ff \Sigma_{\ff t, \ff U}$ and the physical $\ff G=\ff G_{\ff t, \ff U}$ of the system with Hamiltonian $H_{\ff t, \ff U}$:
\be
  \widehat{\ff \Sigma}_{\ff U}[ \ff G_{\ff t, \ff U} ] = \ff \Sigma_{\ff t, \ff U} \: .
\labeq{gex}
\ee
This construction simplifies the original presentation in Ref.\ \cite{Pot06b}. 
The discussion on the existence and the uniqueness of possible solutions of the relation (\ref{eq:rel}) given there applies accordingly to the present case.

With the help of the functionals $\widehat{\Omega}_{\ff U}[\ff G_0^{-1}]$ and $\widehat{\ff \Sigma}_{\ff U}[\ff G]$, the Luttinger-Ward functional is obtained as:
\be
  \widehat{\Phi}_{\ff U}[\ff G] = 
  \widehat{\Omega}_{\ff U}[ 
  \ff G^{-1} + \widehat{\ff \Sigma}_{\ff U}[\ff G]
  ]
  - 
  \Tr \ln \ff G
  +
  \Tr ( \widehat{\ff \Sigma}_{\ff U}[\ff G] \ff G) \: .
\labeq{phidef}
\ee
Let us check property (iv).
Using \refeq{omder} one finds for the derivative of the first term:
\be
  \frac{1}{T} 
  \frac{\delta
  \widehat{\Omega}_{\ff U}[ \ff G^{-1} + \widehat{\ff \Sigma}_{\ff U}[\ff G] ]
  }{\delta G_{\alpha\beta}} 
  =
  - \sum_{\gamma\delta} \widehat{\ff G}_{\ff U,\delta\gamma}[ 
  \ff G^{-1} + \widehat{\ff \Sigma}_{\ff U}[\ff G]
  ] 
  \left(
  \frac{\delta G_{\gamma\delta}^{-1}}{\delta G_{\alpha\beta}} + \frac{\delta \widehat{\Sigma}_{\ff U,\gamma\delta}[\ff G]}{\delta G_{\alpha\beta}} 
  \right)
\ee
and, using \refeq{gfunc},
\be
  \frac{1}{T} 
  \frac{\delta
  \widehat{\Omega}_{\ff U}[ \ff G^{-1} + \widehat{\ff \Sigma}_{\ff U}[\ff G] ]
  }{\delta G_{\alpha\beta}} 
  =
  - \sum_{\gamma\delta} G_{\delta\gamma}
  \left(
  \frac{\delta G_{\gamma\delta}^{-1}}{\delta G_{\alpha\beta}} + \frac{\delta \widehat{\Sigma}_{\ff U,\gamma\delta}[\ff G]}{\delta G_{\alpha\beta}} 
  \right) \: .
\ee
With
\be
  \frac{\delta G^{-1}_{\gamma\delta}}{\delta G_{\alpha\beta}} 
  =
  - G^{-1}_{\gamma\alpha}G^{-1}_{\beta\delta}
\ee
we find:
\be
\frac{1}{T} \frac{\delta \widehat{\Phi}_{\ff U}[\ff G]}{\delta G_{\alpha\beta}} 
  =
  G_{\beta\alpha}^{-1}
  -
  \sum_{\gamma\delta } G_{\delta\gamma} 
  \frac{\delta \widehat{\Sigma}_{\ff U,\gamma\delta}[\ff G]}{\delta G_{\alpha\beta}}
  +
  \frac{1}{T} \frac{\delta}{\delta G_{\alpha\beta}} 
  \left( - 
  \Tr \ln \ff G
  +
  \Tr ( \widehat{\ff \Sigma}_{\ff U}[\ff G] \ff G) 
  \right)
\ee
and thus:
\be
\frac{1}{T} \frac{\delta \widehat{\Phi}_{\ff U}[\ff G]}{\delta G_{\alpha\beta}(i\omega_n)} 
=
\widehat{\Sigma}_{\ff U, \beta\alpha}(i\omega_n)[\ff G] \: ,
\label{eq:pder}
\ee
where, as a reminder, the frequency dependence has been reintroduced.

The other properties are also verified easily. 
(i) and (ii) are obvious.
(iii) follows from \refeq{omex0}, \refeq{calg} and \refeq{gex} and the Dyson equation (\ref{eq:dyson}).
The universality of the Luttinger-Ward functional (v) is ensured by the presented construction. 
It involves universal functionals only.
Finally, (vi) follows from 
$\widehat{\cal \ff G}_{\ff U=0}[\ff G^{-1}] = \ff G$
which implies (via \refeq{gfunc})
$\widehat{\ff \Sigma}_{\ff U=0}[\ff G] = 0$, and with 
$\widehat{\Omega}_{\ff U=0}[\ff G^{-1}] = \Tr \ln \ff G$
we get
$\widehat{\Phi}_{\ff U=0}[\ff G] = 0$.

\subsection{Dynamical variational principle}

The functional $\ff \Sigma_{\ff U}[\ff G]$ can be assumed to be invertible {\em locally} provided that the system is not at a critical point for a phase transition (see also Ref.\ \cite{Pot03a}). 
This allows to construct the Legendre transform of the Luttinger-Ward functional:
\be
  \widehat{F}_{\ff U}[\ff \Sigma] = \widehat{\Phi}_{\ff U}[\widehat{\ff G}_{\ff U}[\ff \Sigma]] 
  - 
  \Tr (\ff \Sigma \widehat{\ff G}_{\ff U}[\ff \Sigma]) \: .
\ee
Here, $\widehat{\ff G}_{\ff U}[\widehat{\ff \Sigma}_{\ff U}[\ff G]] = \ff G$. 
With \refeq{pder} we immediately find
\be
\frac{1}{T} \frac{\delta \widehat{F}_{\ff U}[\ff \Sigma]}{\delta \ff \Sigma} 
=
- \widehat{\ff G}_{\ff U}[\ff \Sigma] \: .
\label{eq:fder}
\ee
We now define the self-energy functional:
\be
  \widehat{\Omega}_{\ff t, \ff U}[\ff \Sigma] = 
  \Tr \ln \frac{1}{\ff G_{\ff t,0}^{-1} - \ff \Sigma}
  + \widehat{F}_{\ff U}[\ff \Sigma] \: .
\labeq{sef}
\ee
Its functional derivative is easily calculated:
\be
  \frac{1}{T} \frac{\delta \widehat{\Omega}_{\ff t, \ff U}[\ff \Sigma]}
  {\delta \ff \Sigma} = 
  \frac{1}{\ff G_{\ff t,0}^{-1} - \ff \Sigma} - 
  \widehat{\ff G}_{\ff U}[\ff \Sigma] \: .
\ee
The equation 
\be
  \widehat{\ff G}_{\ff U}[\ff \Sigma] = \frac{1}{\ff G_{\ff t,0}^{-1} - \ff \Sigma}
\label{eq:sig}
\ee
is a (highly non-linear) conditional equation for the self-energy of the system $H = H_0(\ff t) + H_{1}(\ff U)$.
Eqs.\ (\ref{eq:dyson}) and (\ref{eq:gex}) show that it is satisfied by the physical self-energy $\ff \Sigma = \ff \Sigma_{\ff t, \ff U}$.
Note that the l.h.s of (\ref{eq:sig}) is independent of $\ff t$ but depends on $\ff U$ (universality of $\widehat{\ff G}_{\ff U}[\ff \Sigma]$), while the r.h.s  is independent of $\ff U$ but depends on $\ff t$ via $\ff G_{\ff t,0}^{-1}$.
The obvious problem of finding a solution of \refeq{sig} is that there is no closed form for the functional $\widehat{\ff G}_{\ff U}[\ff \Sigma]$.
Solving Eq.\ (\ref{eq:sig}) is equivalent to a search for the stationary point of the grand potential as a functional of the self-energy:
\be
  \frac{\delta \widehat{\Omega}_{\ff t, \ff U}[\ff \Sigma]}
  {\delta \ff \Sigma} = 0 \; .
\label{eq:var}
\ee
This represents the dynamical variational principle and the starting point for self-energy-functional theory.
It should be mentioned that one can equivalently formulate the principle with the Green's function as the basic variable, i.e.\ $\delta \widehat{\Omega}_{\ff t, \ff U}[\ff G] / \delta \ff G = 0$ (see Refs.\ \cite{CK00,CK01}, for examples). 
One reason to consider self-energy functionals instead of functionals of the Green's function, however, is to derive the dynamical mean-field theory as a type-III approximation.

\subsection{Reference system} 

The central idea of self-energy-functional theory is to make use of the universality of (the Legendre transform of) the Luttinger-Ward functional to construct type-III approximations.
Consider the self-energy functional \refeq{sef}. 
Its first part consists of a simple explicit functional of $\ff \Sigma$ while its second part, i.e.\ $\widehat{F}_{\ff U}[\ff \Sigma]$, is unknown but depends on $\ff U$ only.

Due to this universality of $\widehat{F}_{\ff U}[\ff \Sigma]$, one has
\be
  \widehat{\Omega}_{\ff t', \ff U}[\ff \Sigma] = 
  \Tr \ln \frac{1}{\ff G_{\ff t',0}^{-1} - \ff \Sigma}
  + \widehat{F}_{\ff U}[\ff \Sigma] \: 
\labeq{sfp}
\ee
for the self-energy functional of a so-called ``reference system''. 
As compared to the original system of interest, the reference system is given by a Hamiltonian $H' = H_{0} (\ff t') + H_{1} (\ff U)$ with the same interaction part $\ff U$ but modified one-particle parameters $\ff t'$.
This is different as compared to the static variational principle where reference systems with $\ff U' \ne \ff U$ may be considered.
Note that the reference system has different microscopic parameters but is assumed to be in the 
same macroscopic state, i.e.\ at the same temperature $T$ and the same chemical 
potential $\mu$.
By a proper choice of its one-particle part, the problem posed by the reference system $H'$ can be much simpler than the original problem posed by $H$. 
We assume that the self-energy of the reference system $\ff \Sigma_{\ff t',\ff U}$ can be computed exactly, e.g.\ by some numerical technique.

Combining Eqs.\ (\ref{eq:sef}) and (\ref{eq:sfp}), one can eliminate the unknown functional $\widehat{F}_{\ff U}[\ff \Sigma]$:
\be
  \widehat{\Omega}_{\ff t, \ff U}[\ff \Sigma] 
  = 
  \widehat{\Omega}_{\ff t', \ff U}[\ff \Sigma] 
  + 
  \Tr \ln \frac{1}{\ff G_{\ff t,0}^{-1} - \ff \Sigma}
  - 
  \Tr \ln \frac{1}{\ff G_{\ff t',0}^{-1} - \ff \Sigma} \: .
  \labeq{sfp1}
\ee
It appears that this amounts to a shift of the problem only as the self-energy {\em functional} of the reference system again contains the full complexity of the problem.
In fact, except for the trivial case $\ff U=0$, the functional dependence of $\widehat{\Omega}_{\ff t', \ff U}[\ff \Sigma]$ is unknown -- even if the reference system is assumed to be solvable, i.e.\ if the self-energy $\ff \Sigma_{\ff t',\ff U}$, the Green's function $\ff G_{\ff t',\ff U}$ and the grand potential $\Omega_{\ff t', \ff U}$ of the reference system are available.

However, inserting the self-energy of the reference system $\ff \Sigma_{\ff t',\ff U}$ into the self-energy functional of the original one, and using $\widehat{\Omega}_{\ff t', \ff U}[\ff \Sigma_{\ff t',\ff U}] = \Omega_{\ff t',\ff U}$ and the Dyson equation of the reference system,
we find:
\be
  \widehat{\Omega}_{\ff t, \ff U}[\ff \Sigma_{\ff t',\ff U}] 
  = 
  {\Omega}_{\ff t', \ff U}
  + 
  \Tr \ln \frac{1}{\ff G_{\ff t,0}^{-1} - \ff \Sigma_{\ff t',\ff U}}
  - 
  \Tr \ln \ff G_{\ff t',\ff U} \:  .
\labeq{ocalc}	
\ee
This is a remarkable result.
It shows that an {\em exact} evaluation of the self-energy functional of a non-trivial original system is possible, at least for certain self-energies.
This requires to solve a reference system with the same interaction part.

%*********************************************************************************
\begin{figure}[t]
  \centerline{\includegraphics[width=0.5\textwidth]{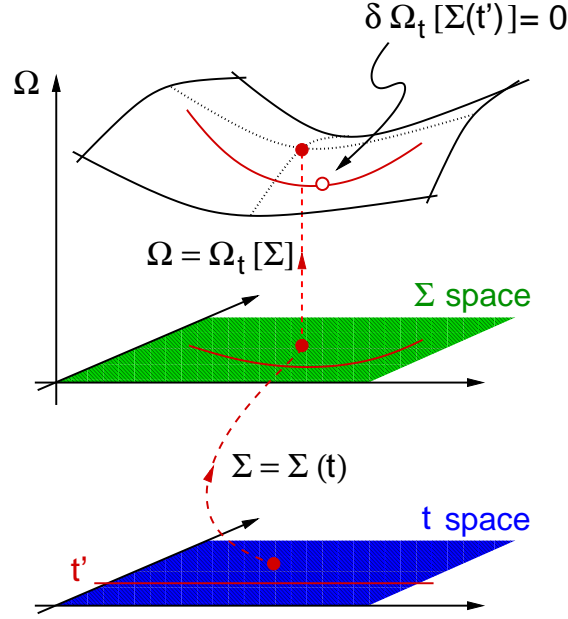}}
\caption{
(taken from Ref.\ \cite{Pot11a}).
Schematic illustration for the construction of consistent approximations within the SFT. 
The grand potential is considered as a functional of the self-energy which is parametrized by the one-particle parameters $\ff t$ of the Hamiltonian ($\ff U$ is fixed).
At the physical self-energy, $\Omega$ is stationary (filled red circles).
The functional dependence of $\Omega$ on $\ff \Sigma$ is not accessible on the entire space of self-energies ($\Sigma$ space) but on a restricted subspace of ``trial'' self-energies parametrized by a subset of one-particle parameters $\ff t'$ (solid red lines) corresponding to a ``reference system'', i.e.\ a manifold of systems with the same interaction part but a one-particle part given by $\ff t'$ which renders the solution possible.
The self-energy functional can be evaluated exactly on the submanifold of reference systems. 
A stationary point on this submanifold represents the approximate self-energy and the corresponding approximate grand potential (open circle).
}
\label{fig:sft}
\end{figure}
%*********************************************************************************

\refeq{ocalc} provides an explicit expression of the self-energy functional $\widehat{\Omega}_{\ff t, \ff U}[\ff \Sigma]$.
This is suitable to discuss the domain of the functional precisely. 
Take $\ff U$ to be fixed.
We define the space of $\ff t$-representable self-energies as 
\be
  {\cal S}_{\ff U} 
  = 
  \{
  \ff \Sigma \: | \: \exists \ff t: \; \ff \Sigma = \ff \Sigma_{\ff t,\ff U}
  \}
  \: .
\ee
This definition of the domain is very convenient since it ensures the correct analytical and causal properties of the variable $\ff \Sigma$.

We can now formulate the above result in the following way. 
Consider a set of reference systems with $\ff U$ fixed but different one-particle parameters $\ff t'$, i.e.\ a space of one-particle parameters ${\cal T'}$. 
Assume that the reference system with $H' = H_{\ff t',\ff U}$ can be solved exactly for any $\ff t' \in {\cal T'}$.
Then, the self-energy functional $\widehat{\Omega}_{\ff t, \ff U}[\ff \Sigma]$ can be evaluated exactly on the subspace 
\be
  {\cal S}'_{\ff U} 
  = 
  \{
  \ff \Sigma \: | \: \exists \ff t' \in {\cal T'}: \; \ff \Sigma = \ff \Sigma_{\ff t',\ff U}
  \}
  \subset 
  {\cal S}_{\ff U} 
  \: .
\ee
This fact can be used to construct type-III approximations, see Fig.\ \ref{fig:sft}.

\subsection{Construction of cluster approximations} 

\begin{figure}[b]
\centering
\includegraphics[width=0.4\columnwidth]{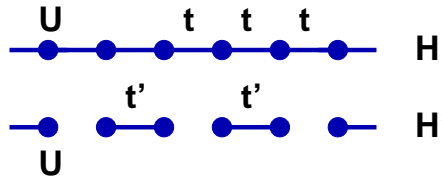}
\caption{
(taken from Ref.\ \cite{Pot11a}).
Variational cluster approximation (VCA) for the Hubbard model. 
Top: representation of the original one-dimensional Hubbard model $H$ with nearest-neighbor hopping $t$ and Hubbard-interaction $U$.
Bottom: reference system $H'$ consisting of decoupled clusters of $L_c=2$ sites each with intra-cluster hopping $t'$ as a variational parameter.
}
\label{fig:refsys}
\end{figure}

A certain approximation is defined by a choice of the reference system or actually by a manifold of reference systems specified by a manifold of one-particle parameters ${\cal T}'$.
As an example consider Fig.\ \ref{fig:refsys}. 
The original system is given by the one-dimensional Hubbard model with nearest-neighbor hopping $t$ and Hubbard interaction $U$.
A possible reference system is given by switching off the hopping between clusters consisting of $L_c=2$ sites each.
The hopping within the cluster $t'$ is arbitrary, this defines the space ${\cal T}'$.
The self-energies in ${\cal S}'$, the corresponding Green's functions and grand potentials of the reference system can obviously be calculated easily since the degrees of freedom are decoupled spatially. 
Inserting these quantities in \refeq{ocalc} yields the SFT grand potential as a function of $\ff t'$:
\be
  \Omega(\ff t') \equiv \widehat{\Omega}_{\ff t, \ff U}[\ff \Sigma_{\ff t',\ff U}] \: .
\labeq{sftgp}
\ee
This is no longer a functional but an ordinary function of the variational parameters $\ff t' \in {\cal T}'$.
The final task then consists in finding a stationary point $\ff t'_{\rm opt}$ of this function:
\be
  \frac{\partial \Omega(\ff t')}{\partial \ff t'} = 0 \qquad \mbox{for} \; \ff t' = \ff t'_{\rm opt} \: .
\labeq{stat}
\ee

\begin{figure}[t]
\centering
\includegraphics[width=0.5\columnwidth]{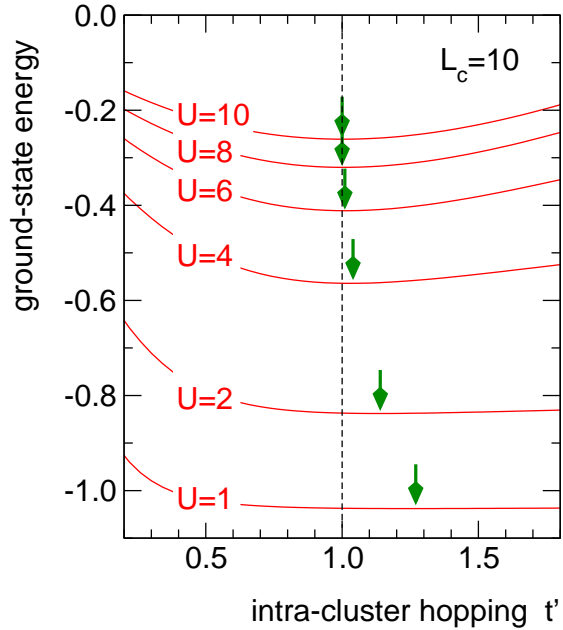}
\caption{
(taken from Ref.\ \cite{BHP08}).
SFT ground-state energy per site, i.e.\ $(\Omega(t')+\mu \langle N\rangle)/L$, as a function of the intra-cluster nearest-neighbor hopping $t'$ for $L_c=10$ and different $U$ ($\mu=U/2$) at zero temperature. 
Arrows indicate the respective optimal $t'$.
The energy scale is fixed by $t=1$.
}
\label{fig:omegat}
\end{figure}

\begin{figure}[t]
\centering
\includegraphics[width=0.6\columnwidth]{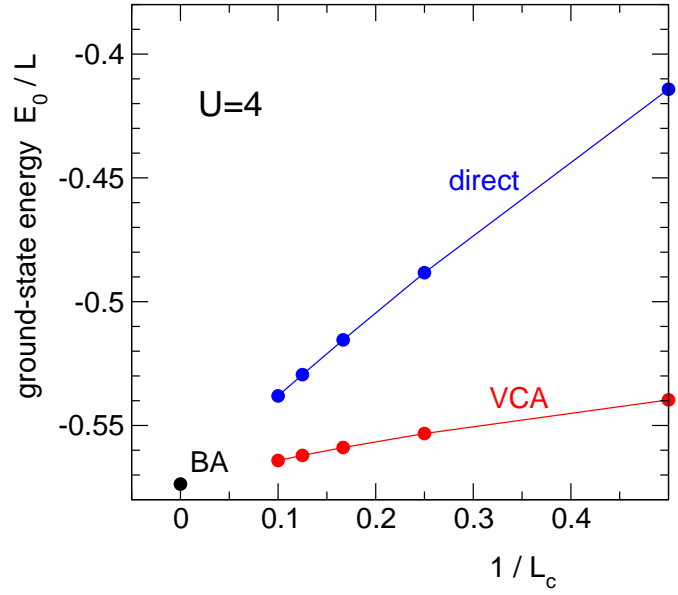}
\caption{
(taken from Ref.\ \cite{BHP08}).
Optimal VCA ground-state energy per site for $U=4$ for different cluster sizes $L_c$ as a function of $1/L_c$ compared to the exact (BA) result and the direct cluster approach.
}
\label{fig:u48}
\end{figure}

\begin{figure}[t]
\centering
\includegraphics[width=0.45\columnwidth]{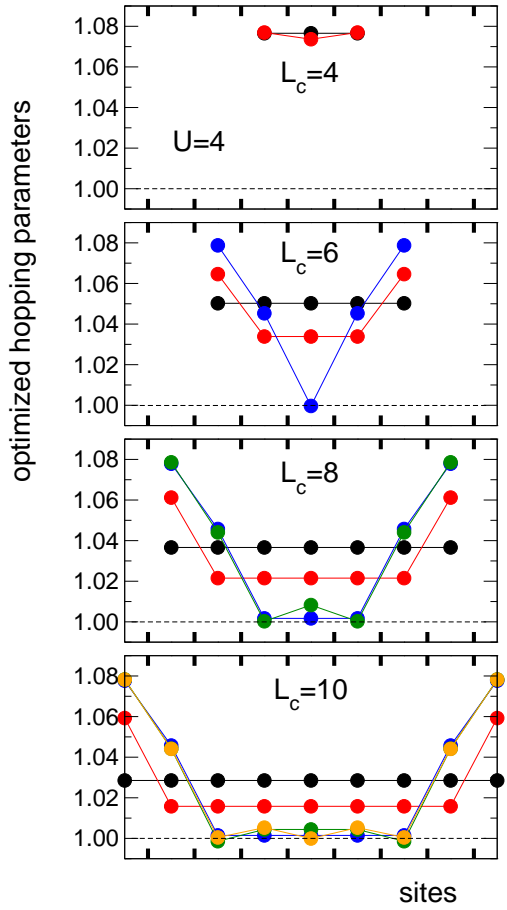}
\caption{
(taken from Ref.\ \cite{BHP08}).
Optimized hopping parameters for the reference systems shown in Fig.\ \ref{fig:refsys} but for larger clusters with $L_c$ sites each as indicated.
VCA results for $U=4$, $t=1$ at half-filling and temperature $T=0$.
{\em Black}: hopping assumed to be uniform.
{\em Red}: two hopping parameters varied independently,
the hopping at the two cluster edges and the ``bulk'' hopping.
{\em Blue}: hopping at the edges, next to the edges and bulk hopping varied.
{\em Green}: four hopping parameters varied.
{\em Orange}: five hopping parameters varied.
}
\label{fig:hopp}
\end{figure}

In the example considered this is a function of a single variable $t'$ (we assume $t'$ to be the same for all clusters).
Note that not only the reference system (in the example the isolated cluster) defines the final result but also the lattice structure and the one-particle parameters of the original system.
These enter $\Omega(\ff t')$ via the free Green's function $G_{\ff t,0}$ of the original system.
In the first term on the r.h.s of \refeq{ocalc} we just recognize the CPT Green's function
$1/(\ff G_{\ff t,0}^{-1} - \ff \Sigma_{\ff t',\ff U})$.
The approximation generated by a reference system of disconnected clusters is called variational cluster approximation (VCA). 

An example for the results of a numerical calculation is given in Fig.\ \ref{fig:omegat}, see also Ref.\ \cite{BHP08}.
The calculation has been performed for the one-dimensional particle-hole symmetric Hubbard model at half-filling and zero temperature.
The figure shows the numerical results for the optimal nearest-neighbor intra-cluster hopping $t'$ as obtained from the VCA for a reference system with disconnected clusters consisting of $L_c=10$ sites each.
The hopping $t'$ is assumed to be the same for all pairs of nearest neighbors.
In principle, one could vary all one-particle parameters that do not lead to a coupling of the clusters to get the optimal result.
In most cases, however, is it necessary to restrict oneself to a small number of physically motivated variational parameters to avoid complications arising from a search for a stationary point in a high-dimensional parameter space.
For the example discussed here, the parameter space ${\cal T}'$ is one-dimensional only.
This is the most simple choice but more elaborate approximations can be generated easily.
The flexibility to construct approximations of different quality and complexity must be seen as one of the advantages of the variational cluster approximation and of the SFT in general.

As can be seen from the figure, a non-trivial result, namely $t'_{\rm opt} \ne t=1$, is found as the optimal value of $t'_{\rm opt}$. 
We also notice that $t_{\rm opt} > t$. 
The physical interpretation is that switching off the {\em inter}-cluster hopping, which generates the approximate self-energy, can partially be compensated for by enhancing the {\em intra}-cluster hopping.
The effect is the more pronounced the smaller is the cluster size $L_c$.
Furthermore, it is reasonable that in case of a stronger interaction and thus more localized electrons, switching off the inter-cluster hopping is less significant. 
This can be seen in Fig.\ \ref{fig:omegat}: 
The largest optimal hopping $t'_{\rm opt}$ is obtained for the smallest $U$.

On the other hand, even a ``strong'' approximation for the self-energy (measured as a strong deviation of $t'_{\rm opt}$ from $t$) becomes irrelevant in the weak-coupling limit because the self-energy must vanish for $U=0$.
Generally, we note that the VCA becomes exact in the limit $\ff U = 0$: In \refeq{sfp1} the first and the third terms on r.h.s cancel each other and we are left with 
\be
  \widehat{\Omega}_{\ff t, \ff U=0}[\ff \Sigma] 
  = 
  \Tr \ln \frac{1}{\ff G_{\ff t,0}^{-1} - \ff \Sigma} \; .
  \labeq{sfp10}
\ee
Since the trial self-energy has to be taken from a reference system with the same interaction part, i.e.\ $\ff U=0$, we have $\ff \Sigma=0$, and the limit becomes trivial.
For weak but finite $\ff U$, the SFT grand potential becomes flatter and flatter until for $\ff U=0$ the $\ff t'$ dependence is completely irrelevant.

The VCA is also exact in the atomic limit or, more general but again trivial, in the case that there is no restriction on the trial self-energies: ${\cal S}' = {\cal S}$.
In this case, $\ff t'_{\rm opt} = \ff t$ solves the problem, i.e.\ the second and the third term on the r.h.s of \refeq{ocalc} cancel each other and $\widehat{\Omega}_{\ff t, \ff U}[\ff \Sigma_{\ff t',\ff U}] = {\Omega}_{\ff t', \ff U}$ for $\ff t' = \ff t$.

Cluster-perturbation theory (CPT) can be understood as being identical with the VCA provided that the SFT expression for the grand potential is used and that no parameter optimization at all is performed. 
As can be seen from Fig.\ \ref{fig:omegat}, there is a gain in binding energy due to the optimization of $t'$, i.e.\ $\Omega(t'_{\rm opt}) < \Omega(t)$.
This means that the VCA improves on the CPT result.

Fig.\ \ref{fig:u48} shows the ground-state energy (per site), i.e.\ the SFT grand potential at zero temperature constantly shifted by $\mu N$ at the stationary point, as a function of the inverse cluster size $1/L_c$.
The dependence turns out to be quite regular and allows to recover the exact Bethe-Ansatz result (BA) \cite{LW68} by extrapolation to $1/L_c = 0$. 
The VCA result represents a considerable improvement as compared to the ``direct'' cluster approach where $E_0$ is simply approximated by the ground-state energy of an isolated Hubbard chain (with open boundary conditions).
Convergence to the exact result is clearly faster within the VCA. 
Note that the direct cluster approach, opposed to the VCA, is not exact for $U=0$.

In the example discussed so far a single variational parameter was taken into account only.
More parameters can be useful for different reasons. 
For example, the optimal self-energy provided by the VCA as a real-space cluster technique artificially breaks the translational symmetry of the original lattice problem. 
Finite-size effects are expected to be the most pronounced at the cluster boundary.
This suggests to consider all intra-cluster hopping parameters as independent variational parameters or at least the hopping at the edges of the chain.

The result is shown in Fig.\ \ref{fig:hopp}.
We find that the optimal hopping varies between different nearest neighbors within a range of less than 10\%. 
At the chain edges the optimal hopping is enhanced to compensate the loss of itinerancy due to the switched-off inter-cluster hopping within the VCA.
With increasing distance to the edges, the hopping quickly decreases.
Quite generally, the third hopping parameter is already close to the physical hopping $t$. 
Looking at the $L_c=10$ results where all (five) different hopping parameters have been varied independently (orange circles), one can see the hopping to slightly oscillate around the bulk value reminiscent of surface Friedel oscillations. 

The optimal SFT grand potential is found to be lower for the inhomogeneous cases as compared to the homogeneous (black) one.
Generally, the more variational parameters are taken into account the higher is the decrease
of the SFT grand potential at optimal parameters.
However, the binding-energy gain due to inhomogeneous hopping parameters is much smaller compared to the gain obtained with a larger cluster. 

Considering an additional hopping parameter $t_{\rm pbc}$ linking the two chain edges as a variational parameter, i.e.\ clusters with periodic boundary conditions always gives a minimal SFT grand potential at $t_{\rm pbc}=0$
(instead of a stationary point at $t_{\rm pbc} = 1$). 
This implies that open boundary conditions are preferred (see also Ref.\ \cite{PAD03}).

\section{Consistency, symmetry and systematics}
\label{sec:8}

\subsection{Analytical structure of the Green's function}

Constructing approximations within the framework of a dynamical variational principle means that, besides an approximate thermodynamical potential, approximate expressions for the self-energy and the one-particle Greens function are obtained.
This raises the question whether their correct analytical structure is respected in an approximation.
For approximations obtained from self-energy-functional theory this is easily shown to be the case in fact.

The physical self-energy $\Sigma_{\alpha\beta}(\omega)$ and the physical Green's function $G_{\alpha\beta}(\omega)$ are analytical functions in the entire complex $\omega$ plane except for the real axis and have a spectral representation (see \refeq{spectralrep}) with non-negative diagonal elements of the spectral function. 

This trivially holds for the SFT self-energy $\Sigma_{\ff t',\ff U, \alpha\beta}(\omega)$ since by construction $\Sigma_{\ff t',\ff U, \alpha\beta}(\omega)$ is the {\em exact} self-energy of a reference system.
The SFT Green's function is obtained from the SFT self-energy and the free Green's function of the original model via Dyson's equation: 
\be
\ff G = \frac{1}{\ff G_{\ff t,0}^{-1} - \ff \Sigma_{\ff t',\ff U}} \: .
\ee
It is easy to see that it is analytical in the complex plane except for the real axis.
To verify that is has a spectral representation with non-negative spectral function, we can equivalently consider the corresponding retarded quantity
$\ff G_{{\rm ret}} (\omega) = \ff G(\omega + i 0^+)$ for real frequencies $\omega$
and verify that ${\bm G}_{{\rm ret}} = {\bm G}_{\rm R} - i {\bm G}_{\rm I}$ with 
${\bm G}_{\rm R}$, ${\bm G}_{\rm I}$ Hermitian and ${\bm G}_{\rm I}$ non-negative, i.e.\ $\ff x^\dagger \ff G_{\rm I} \ff x \ge 0$ for all $\ff x^\dagger = (\dots,x_\alpha,\dots)$.

This is shown as follows:
We can assume that 
${\bm G}_{0,{\rm ret}} = {\bm G}_{0, \rm R} - i {\bm G}_{0, \rm I}$ with 
${\bm G}_{0, \rm R}$, ${\bm G}_{0, \rm I}$ Hermitian and 
${\bm G}_{0, \rm I}$ non-negative.
Since for Hermitian matrices ${\bm A}$, ${\bm B}$ with ${\bm B}$ non-negative, one has
$1/({\bm A} \pm i {\bm B}) = {\bm X} \mp i {\bm Y}$
with ${\bm X}$, ${\bm Y}$ Hermitian and ${\bm Y}$ non-negative (see Ref.\ \cite{Pot03b}), 
we find
${\bm G}_{0,{\rm ret}}^{-1} = {\bm P}_{\rm R} + i {\bm P}_{\rm I}$
with ${\bm P}_{\rm R}$, ${\bm P}_{\rm I}$ Hermitian and ${\bm P}_{\rm I}$ 
non-negative.
Furthermore, we have 
${\bm \Sigma}_{\rm ret} = {\bm \Sigma}_{\rm R} - i {\bm \Sigma}_{\rm I}$ 
with ${\bm \Sigma}_{\rm R}$, ${\bm \Sigma}_{\rm I}$ Hermitian and 
${\bm \Sigma}_{\rm I}$ non-negative. 
Therefore,
\begin{equation}
  {\bm G}_{\rm ret} 
  = \frac{1}{{\bm P}_{\rm R} + i {\bm P}_{\rm I} 
  - {\bm \Sigma}_{\rm R} + i {\bm \Sigma}_{\rm I}}
  = \frac{1}{{\bm Q}_{\rm R} + i {\bm Q}_{\rm I}}
\end{equation}
with ${\bm Q}_{\rm R}$ Hermitian and ${\bm Q}_{\rm I}$ Hermitian 
and non-negative.

Note that the proof given here is simpler than corresponding proofs for cluster extensions of the DMFT \cite{HMJK00,KSPB01} because the SFT does not involve a ``self-consistency condition'' which is the main object of concern for possible causality violations.

\subsection{Thermodynamical consistency} 

An advantageous feature of the VCA and of other approximations within the SFT framework is their internal thermodynamical consistency. 
This is due to the fact that all quantities of interest are derived from an approximate but explicit expression for a thermodynamical potential. 
In principle the expectation value of any observable should be calculated by via
\be
  \langle A \rangle = \frac{\partial \Omega}{\partial \lambda_A} \; ,
\ee
where $\Omega \equiv \Omega(\ff t')$ is the SFT grand potential (see \refeq{sftgp}) at $\ff t' = \ff t'_{\rm opt}$ and $\lambda_A$ is a parameter in the Hamiltonian of the original system which couples linearly to $A$, i.e.\ $H_{\ff t,\ff U} = H_0 + \lambda_A A$.
This ensures, for example, that the Maxwell relations
\be
  \frac{\partial \langle A \rangle}{\partial \lambda_B} 
  = 
  \frac{\partial \langle B \rangle}{\partial \lambda_A} 
\ee
are respected. 

Furthermore, thermodynamical consistency means that expectation values of arbitrary one-particle operators $A= \sum_{\alpha\beta} A_{\alpha\beta} c^\dagger_\alpha c_\beta$ can consistently either be calculated by a corresponding partial derivative of the grand potential on the one hand, or by integration of the one-particle spectral function on the other. 
As an example we consider the total particle number $N=\sum_\alpha c_\alpha^\dagger c_\alpha$. 
{\em A priori} it not guaranteed that in an approximate theory the expressions
\begin{equation}
  \langle N \rangle = - \frac{\partial \Omega}{\partial \mu} \: ,
\label{eq:n1}
\end{equation}  
and
\begin{equation}
  \langle N \rangle = \sum_{\alpha} \int_{-\infty}^\infty f(z) A_{\alpha\alpha}(z) dz \: 
\label{eq:n2}
\end{equation}  
with $f(z) = 1/ (\exp(\beta z) + 1)$ and the spectral function $A_{\alpha\beta}(z)$ will give the same result.

To prove thermodynamic consistency, we start from Eq.\ (\ref{eq:n1}).
According to \refeq{ocalc}, there is a twofold $\mu$ dependence of $\Omega \equiv \Omega_{\ff t, \ff U} [\ff \Sigma_{\ff t'_{\rm opt},\ff U}]$:
(i) the {\em explicit} $\mu$ dependence due to the chemical potential in the free Green's function of the original model, $\ff G_{\ff t,0}^{-1} = \omega + \mu - \ff t$,
and (ii) an {\em implicit} $\mu$ dependence due to the $\mu$ dependence of the self-energy $\ff \Sigma_{\ff t'_{\rm opt},\ff U}$, the Green's function $\ff G_{\ff t'_{\rm opt},\ff U}$ and the grand potential $\Omega_{\ff t'_{\rm opt},\ff U}$ of the reference system:
\begin{equation}
  \langle N \rangle 
  = 
  - \frac{\partial \Omega}{\partial \mu_{\rm ex.}} 
  - \frac{\partial \Omega}{\partial \mu_{\rm im.}} \: .
\labeq{exim}
\end{equation}
Note that the implicit $\mu$ dependence is due to the chemical potential of the reference system which, by construction, is in the same macroscopic state as the original system {\em as well as} due to the $\mu$ dependence of the stationary point $\ff t'_{\rm opt}$ itself.
The latter, however, can be ignored since
\begin{equation}
  \frac{\partial \Omega}{\partial \ff t'} 
  \cdot
  \frac{\partial \ff t'}{\partial \mu} = 0
\end{equation}
for $\ff t' = \ff t'_{\rm opt}$ because of stationarity condition \refeq{stat}.
 
We assume that an overall shift of the one-particle energies $\varepsilon' \equiv t'_{\alpha\alpha}$ is included in the set ${\cal T}'$ of variational parameters.
Apart from the sign this is essentially the ``chemical potential'' in the reference system but should be formally distinguished from $\mu$ since the latter has a macroscopic thermodynamical meaning and is the same as the chemical potential of the original system which should not be seen as a variational parameter.

The self-energy and the Green's function of the reference system are defined as grand-canonical averages.
Hence, their $\mu$ dependence due to the grand-canonical Hamiltonian ${\cal H'} = H' - \mu N$ is (apart from the sign) the same as their dependence on $\varepsilon'$:
Consequently, we have:
\begin{equation}
  \frac{\partial \Omega}{\partial \mu_{\rm im.}} 
  =
  - \frac{\partial \Omega}{\partial \varepsilon'}
  = 0
\end{equation}
due to the stationarity condition again.

We are then left with the explicit $\mu$ dependence only:
\be
  \frac{\partial \Omega}{\partial \mu_{\rm ex.}} 
  =
  \frac{\partial}{\partial \mu_{\rm ex.}}
  \Tr \ln \frac{1}{ \ff G_{\ff t,0}^{-1} - \ff \Sigma_{\ff t'_{\rm opt},\ff U}}
  =
  - \Tr \frac{1}{ \ff G_{\ff t,0}^{-1} - \ff \Sigma_{\ff t'_{\rm opt},\ff U}} \; .
\ee
Converting the sum over the Matsubara frequencies implicit in the trace $\Tr$ into a contour integral in the complex $\omega$ plane and using Cauchy's theorem, we can proceed to an integration over real frequencies.
Inserting into Eq.\ (\ref{eq:exim}), this yields:
\begin{equation}
  \langle N \rangle 
   = 
  - \frac{1}{\pi} \mbox{Im}
  \int_{-\infty}^\infty f(\omega) \: \tr
  \frac{1}{\ff G_{0,\ff t}^{-1}
  - \ff \Sigma_{\ff t',\ff U}} \Bigg|_{\omega + i 0^+} d\omega 
\label{eq:nn}  
\end{equation}
for $\ff t' = \ff t'_{\rm opt}$ which is just the average particle number given by (\ref{eq:n2}). 
This completes the proof. 

\subsection{Symmetry breaking}

The above discussion has shown that besides intra-cluster hopping parameters it can also be important to treat one-particle energies in the reference cluster as variational parameters. 
In particular, one may consider variational parameters which lead to a lower symmetry of the Hamiltonian. 

As an example consider the Hubbard model on a bipartite lattice as the system of interest and disconnected clusters of size $L_c$ as a reference system. 
The reference-system Hamiltonian shall include, e.g., an additional a staggered magnetic-field term:
\be
  H'_{\rm fict.} = B' \sum_{i\sigma} z_i (n_{i\uparrow} - n_{i\downarrow}) \; ,
\ee
where $z_i = +1$ for sites on sublattice 1, and $z_i = -1$ for sublattice 2. 
The additional term $H'_{\rm fict.}$ leads to a valid reference system as there is no change of the interaction part.
We include the field strength $B'$ in the set of variational parameters, $B' \in {\cal T}'$.

$B'$ is the strength of a fictitious field or, in the language of mean-field theory, the strength of the internal magnetic field or the Weiss field.
This has to be distinguished clearly from an external {\em physical} field applied to the system with field strength $B$:
\be
  H_{\rm phys.} = B \sum_{i\sigma} z_i (n_{i\uparrow} - n_{i\downarrow}) \; 
\ee
This term adds to the Hamiltonian of the original system.

We expect $B'_{\rm opt}=0$ in case of the paramagnetic state and $B=0$ (and this is easily verified numerically). 
Consider the $B'$ and $B$ dependence of the SFT grand potential $\Omega(B',B) = \Omega_B[\ff \Sigma_{B'}]$. 
Here we have suppressed the dependencies on other variational parameters $\ff t'$ and on $\ff t,\ff U$.
Due to the stationarity condition, $\partial \Omega(B',B) / \partial B' = 0$, the optimal Weiss field $B'$ can be considered as a function of $B$, i.e.\ $B'_{\rm opt}=B'(B)$. 
Therefore, we also have:
\be
\frac{d}{dB} \frac{\partial \Omega(B'(B),B)}{\partial B'} = 0 \: .
\labeq{cccond}
\ee
This yields:
\be
\frac{\partial^2 \Omega(B'(B),B)}{\partial {B'}^2}
\frac{d B'(B)}{dB} 
+
\frac{\partial^2 \Omega(B'(B),B)}{\partial B \partial B'}
= 0 \: .
\labeq{cccond1}
\ee
Solving for $dB'/dB$ we find:
\be
\frac{dB'}{dB} 
= -
\left[\frac{\partial^2 \Omega}{\partial {B'}^2}\right]^{-1}
\frac{\partial^2 \Omega}{\partial B \partial B'}
\: .
\labeq{cccond2}
\ee
This clearly shows that $B'=B'_{\rm opt}$ has to be interpreted carefully. 
$B'$ can be much stronger than $B$ if the curvature $\partial^2 \Omega/\partial {B'}^2$ of the SFT functional at the stationary point is small, i.e.\ if the functional is rather flat as it is the case in the limit $U\to 0$, for example.

From the SFT approximation for the staggered magnetization, 
\be
  m = \sum_{i\sigma} z_i \langle (n_{i\uparrow} - n_{i\downarrow}) \rangle
     \approx \frac{d}{dB}\Omega(B'(B),B) 
     = \frac{\partial \Omega(B'(B),B) }{\partial B} \; ,
\ee
where the stationarity condition has been used once more, 
we can calculate the susceptibility,
\be
\chi = \frac{dm}{dB} 
= 
\frac{\partial^2 \Omega(B'(B),B)}{\partial B' \partial B} 
\frac{d B'(B)}{dB} 
+
\frac{\partial^2 \Omega(B'(B),B)}{\partial B^2}  \: .
\ee
Using \refeq{cccond2},
\be
\chi
= 
\frac{\partial^2 \Omega}{\partial B^2} 
-
\left( \frac{\partial^2 \Omega}{\partial {B'}^2} \right)^{-1}
\:
\left( \frac{\partial^2 \Omega}{\partial B' \partial B} \right)^2 
\: .
\labeq{chi}
\ee
We see that there are two contributions. 
The first term is due to the explicit $B$ dependence in the SFT grand potential while the second is due to the implicit $B$ dependence via the $B$ dependence of the stationary point.

\refeq{chi} also demonstrates (see Ref.\ \cite{Ede09}) that for the calculation of the paramagnetic susceptibility $\chi$ one may first consider spin-independent variational parameters only to find a stationary point.
This strongly reduces the computational effort \cite{BP10}.
Once a stationary point is found, partial derivatives according to \refeq{chi} have to calculated with spin-dependent parameters in a single final step.

\subsection{Spontaneous symmetry breaking}

\begin{figure}[b]
\centering
\includegraphics[width=0.6\columnwidth]{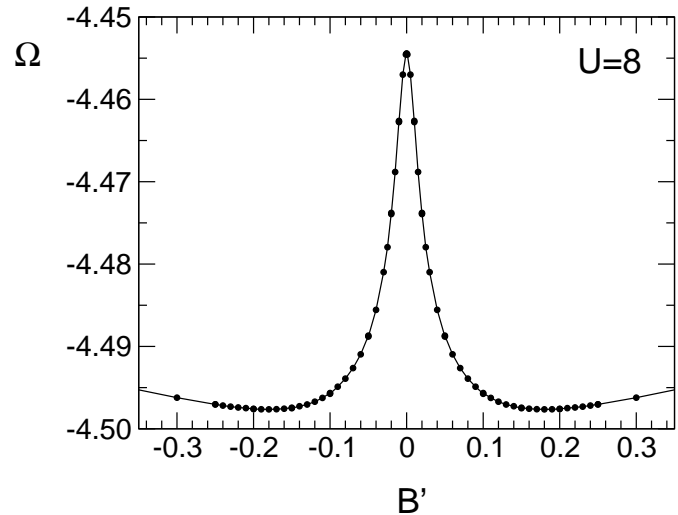}
\caption{
(taken from Ref.\ \cite{DAH+04}).
SFT grand potential as a function of the strength of a fictitious staggered magnetic field $B'$.
VCA calculation using disconnected clusters consisting of $L_c=10$ sites each for the two-dimensional Hubbard model on the square lattice at half-filling, zero temperature, $U=8$ and nearest-neighbor hopping $t=1$.
}
\label{fig:u8}
\end{figure}

\begin{figure}[t]
\centering
\includegraphics[width=0.7\columnwidth]{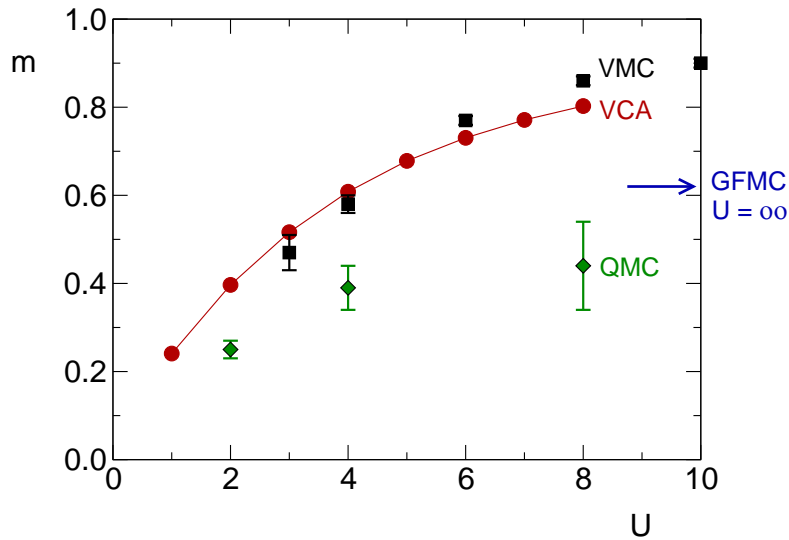}
\caption{
(taken from Ref.\ \cite{DAH+04}).
Comparison of the staggered magnetization $m$ as a function of $U$ at 
half-filling obtained by different methods:
variational cluster approximation (VCA), variational Monte-Carlo (VMC) \cite{YS87} 
and quantum Monte-Carlo (QMC) \cite{HT89}, see text.
The arrow indicates the result $m=0.62 \pm 0.04$ of a Green's-function 
Monte-Carlo study \cite{TC89} for the two-dimensional Heisenberg model.
}
\label{fig:smag}
\end{figure}

{\em Spontaneous} symmetry breaking is obtained at $B=0$ if there is a stationary point with $B'_{\rm opt} \ne 0$. 
Fig.\ \ref{fig:u8} gives an example for the particle-hole symmetric two-dimensional Hubbard model on the square lattice at half-filling and zero temperature. 
As a reference system a cluster with $L_c=10$ sites is considered, and the fictitious staggered magnetic field is taken as the only variational parameter.
There is a stationary point at $B'=0$ which corresponds to the paramagnetic phase. 
At $B'=0$ the usual cluster-perturbation theory is recovered.
The two equivalent stationary points at finite $B'$ correspond to a phase with spontaneous antiferromagnetic order. 
As can be seen from the figure, the antiferromagnetic ground state is stable as compared to the paramagnetic phase.

Its order parameter, i.e.\ the staggered magnetization $m$, is the conjugate variable to the physical field. 
Since the corresponding fictitious field is a variational parameter, $m$ can either be calculated by integration of the spin-dependent spectral density or as the derivative of the SFT grand potential with respect to the physical field strength $B$ with the same result. 

Fig.\ \ref{fig:smag} shows the $U$ dependence of $m$ as calculated at the respective stable stationary point. 
As expected for the parameter regime considered here, a magnetically ordered ground state is obtained for any $U>0$. 
The order parameter monotonously increases with increasing $U$ and reaches a value close to $m=1$ in the $U\to \infty$ limit. 
The trend agrees well with a variational Monte-Carlo calculation \cite{YS87} which is a static mean-field approach based on the Ritz principle but includes correlation effects beyond the Hartree-Fock approximation by partially projecting out doubly occupied sites from a symmetry-broken trial wave function.
It disagrees, however, with the result of auxiliary-field quantum Monte-Carlo calculations \cite{HT89} for finite lattices. 
Within the QMC, the order parameter is obtained from simulations of the static spin-spin correlation function at low temperatures.
It is assumed that the system effectively behaves as if at $T=0$ when the thermal correlation length exceeds the cluster dimensions. 
The VMC and QMC data displayed in Fig.\ \ref{fig:smag} are extrapolated to infinite lattice size.

In the $U\to \infty$ limit where the low-energy sector of the Hubbard model at half-filling can be mapped onto an antiferromagnetic Heisenberg model, the order parameter as obtained within the VCA is overestimated. 
Physically, one would expect a reduction of the staggered magnetization from its saturated value $m=1$ due to non-local transverse spin-spin correlations \cite{Man91} as it is shown, for example, by a Greens's-function Monte-Carlo study \cite{TC89} for the two-dimensional Heisenberg model (see arrow in Fig.\ \ref{fig:smag}). 

The VCA exactly takes into account electron correlations on a length scale given by the linear size of the reference cluster. 
The example shows, however, that longer-ranged correlations are neglected and that the approximation is like a static mean-field approximation beyond the cluster extension. 
It should be regarded as a cluster mean-field approach which is applicable if the physical properties are dominated by short-range correlations on a scale accessible by an exactly solvable finite cluster.

The possibility to study spontaneous symmetry breaking using the VCA with suitably chosen Weiss fields as variational parameters has been exploited frequently in the past. 
Besides antiferromagnetism \cite{DAH+04,NSST08,HKSO08,YO09}, spiral phases \cite{SS08}, ferromagnetism \cite{BP10}, $d$-wave superconductivity \cite{SLMT05,AA05,AAPH06a,AAPH06b,SS06,AADH09,SS09}, charge order \cite{AEvdLP+04,ASE05} and orbital order \cite{LA09} have been investigated. 
The fact that an explicit expression for a thermodynamical potential is available allows to study discontinuous transitions and phase separation as well.

\subsection{Non-perturbative conserving approximations}

Continuous symmetries of a Hamiltonian imply the existence of conserved quantities:
The conservation of total energy, momentum, angular momentum, spin and particle number is enforced by a not explicitly time-dependent Hamiltonian which is spatially homogeneous and isotropic and invariant under global SU(2) and U(1) gauge transformations.
Approximations may artificially break symmetries and thus lead to unphysical violations of conservations laws.
Baym and Kadanoff \cite{BK61,Bay62} have analyzed under which circumstances an approximation respects the mentioned macroscopic conservation laws.
Within diagrammatic perturbation theory it could be shown that approximations which derive from an explicit but approximate expression for the Luttinger-Ward functional $\Phi$ (''$\Phi$-derivable approximations'') are ``conserving''.
Examples for conserving approximations are the Hartree-Fock or the fluctuation-exchange approximation \cite{BK61,BSW89}.

The SFT provides a framework to construct $\Phi$-derivable approximations for correlated lattice models which are {\em non-perturbative}, i.e.\ do not employ truncations of the skeleton-diagram expansion.
Like in weak-coupling conserving approximations, approximations within the SFT are derived from the Luttinger-Ward functional, or its Legendre transform $F_{\ff U}[\ff \Sigma]$. 
These are $\Phi$-derivable since any type-III approximation can also be seen as a type-II one, see Section \ref{sec:types}. 

For fermionic lattice models, conservation of energy, particle number and spin have to be considered.
Besides the static thermodynamics, the SFT concentrates on the {\em one-particle} excitations. 
For the approximate one-particle Green's function, however, it is actually simple to prove directly that the above conservation laws are respected.
A short discussion is given in Ref.\ \cite{OBP07}.

At zero temperature $T=0$ there is another non-trivial theorem which is satisfied by any $\Phi$-derivable approximation, namely Luttinger's sum rule \cite{LW60,Lut60}.
This states that at zero temperature the volume in reciprocal space that is enclosed by the Fermi surface is equal to the average particle number.
The original proof of the sum rule by Luttinger and Ward \cite{LW60} is based on the skeleton-diagram expansion of $\Phi$ in the exact theory and is straightforwardly transferred to the case of a $\Phi$-derivable approximation. 
This also implies that other Fermi-liquid properties, such as the linear trend of the specific heat at low $T$ and Fermi-liquid expressions for the $T=0$ charge and the spin susceptibility are respected by a $\Phi$-derivable approximation.

For approximations constructed within the SFT, a different proof has to be found.
One can start with \refeq{ocalc} and perform the zero-temperature limit for an original system (and thus for a reference system) of finite size $L$. 
The different terms in the SFT grand potential then consist of finite sums.
The calculation proceeds by taking the $\mu$-derivative, for $T=0$, on both sides of \refeq{ocalc}. 
This yields the following result (see Ref.\ \cite{OBP07} for details):
\be
  \langle N \rangle = \langle N \rangle' + 2 \sum_{\ff k} \Theta(G_{\ff k}(0))
  - 2 \sum_{\ff k} \Theta(G'_{\ff k}(0)) \: .
\labeq{res}
\ee
Here $\langle N \rangle$ ($\langle N \rangle'$) is the ground-state expectation value of the total particle number $N$ in the original (reference) system, and $G_{\ff k}(0)$ ($G'_{\ff k}(0)$) are the diagonal elements of the one-electron Green's function $\ff G$ at $\omega=0$.
As Luttinger's sum rule reads
\be
  \langle N \rangle = 2 \sum_{\ff k} \Theta(G'_{\ff k}(0)) \: ,
\ee
this implies that, within an approximation constructed within the SFT, the sum rule is satisfied if and only if it is satisfied for the reference system, i.e.\ if $\langle N \rangle'=2 \sum_{\ff k} \Theta(G'_{\ff k}(0))$.
This demonstrates that the theorem is ``propagated'' to the original system irrespective of the approximation that is constructed within the SFT.
This propagation also works in the opposite direction. 
Namely, a possible violation of the exact sum rule for the reference system would 
imply a violation of the sum rule, expressed in terms of approximate quantities, 
for the original system. 

There are no problems to take the thermodynamic limit $L\to \infty$ (if desired) on both sides of \refeq{res}.
The $\ff k$ sums turn into integrals over the unit cell of the reciprocal lattice.
For a $D$-dimensional lattice the $D-1$-dimensional manifold of $\ff k$ points with 
$G_{\ff k}(0)=\infty$ or $G_{\ff k}(0)=0$ form Fermi or Luttinger surfaces, respectively.
Translational symmetry of the original as well as the reference system may be assumed but is not necessary.
In the absence of translational symmetry, however, one has to re-interpret the wave vector $\ff k$ as an index which refers to the elements of the diagonalized Green's function matrix $\ff G$.
The exact sum rule generalizes accordingly but can no longer be expressed in terms of a Fermi surface since there is no reciprocal space. 
It is also valid for the case of a translationally symmetric original Hamiltonian where, due to the choice of a reference system with reduced translational symmetries, such as employed in the VCA, the symmetries of the (approximate) Green's function of the original system are (artificially) reduced.
Since with \refeq{res} the proof of the sum rule is actually shifted to the proof of the sum rule for the reference system only, we are faced with the interesting problem of the validity of the sum rule for a finite cluster.
For small Hubbard clusters with non-degenerate ground state this has been checked numerically with the surprising result that violations of the sum rule appear in certain parameter regimes close to half-filling, see Ref.\ \cite{OBP07}.
This leaves us with the question where the proof of the sum rule fails if applied to a system of finite size.
This is an open problem that has been stated and discussed in Refs.\ \cite{OBP07,KP07} and that is probably related to the breakdown of the sum rule for Mott insulators \cite{Ros07}. 

\subsection{Systematics of approximations}

Since the SFT unifies different dynamical approximations within a single formal framework, the question arises how to judge on the relative quality of two different approximations resulting from two different reference systems.
This, however, is not straightforward for several reasons.
First, it is important to note that a stationary point of the self-energy functional is not necessarily a minimum but rather a saddle point in general (see Ref.\ \cite{Pot03a} for an example).
The self-energy functional is not convex.
Actually, despite several recent efforts \cite{Kot99,CK01,NST08}, there is no functional relationship between a thermodynamical potential and time-dependent correlation functions, Green's functions, self-energies, etc.\ which is known to be convex.

Furthermore, there is no {\em a priori} reason why, for a given reference system, the SFT grand potential at a stationary point should be lower than the SFT grand potential at another one that results from a simpler reference system, e.g.\ a smaller cluster.
This implies that the SFT does not provide upper bounds to the physical grand potential. 
There is e.g.\ no proof (but also no counterexample) that the DMFT ground-state energy at zero temperature must be higher than the exact one. 
On the other hand, in practical calculations the upper-bound property is usually found to be respected, as can be seen for the VCA in Fig.\ \ref{fig:u48}, for example.
Nevertheless, the non-convexity must be seen as a disadvantage as compared to methods based on wave functions which via the Ritz variational principle are able to provide strict upper bounds.

To discuss how to compare two approximations within SFT, we first have to distinguish between ``trivial'' and ``non-trivial'' stationary points for a given reference system. 
A stationary point is referred to as ``trivial'' if the one-particle parameters are such that the reference system decouples into smaller subsystems.
If, at a stationary point, all degrees of freedom (sites) are still coupled to each other, the stationary point is called ``non-trivial''.
It is possible to prove the following theorem \cite{Pot06a}:
Consider a reference system with a set of variational parameters $\ff t' = \ff t'' + \ff V$ where $\ff V$ couples two separate subsystems.
For example, $\ff V$ could be the inter-cluster hopping between two subclusters with completely decouples the degrees of freedom for $\ff V=0$ and all $\ff t''$.
Then,
\be
  \Omega_{\ff t,\ff U}[\ff \Sigma_{\ff t'' + \ff V}] = \Omega_{\ff t,\ff U}[\ff \Sigma_{\ff t''+0}] + {\cal O}(\ff V^2) \; ,
\ee
provided that the functional is stationary at $\ff \Sigma_{\ff t'',\ff U}$ {\em when varying $\ff t''$ only}  (this restriction makes the theorem non-trivial).
This means that going from a more simple reference system to a more complicated one with more degrees of freedom coupled, should generate a new non-trivial stationary point with $\ff V\ne 0$ while the ``old'' stationary point with $\ff V=0$ being still a stationary point with respect to the ``new'' reference system. 
Coupling more and more degrees of freedom introduces more and more stationary points, and none of the ``old'' ones is ``lost''.

Consider a given reference system with a non-trivial stationary point and a number of trivial stationary points.
An intuitive strategy to decide between two stationary points would be to always take the one with the lower grand potential $\Omega_{\ff t,\ff U}[\ff \Sigma_{\ff t',\ff U}]$.
A sequence of reference systems (e.g. $H'_{\rm A}$, $H'_{\rm B}$, $H'_{\rm C}$, ...) in which more and more degrees of freedom are coupled and which eventually recovers the original system $H$ itself, shall be called a ``systematic'' sequence of reference systems.
For such a systematic sequence, the suggested strategy trivially produces a series of stationary points with monotonously decreasing grand potential.
Unfortunately, however, the strategy is useless because it cannot ensure that a systematic sequence of reference systems generates a systematic sequence of approximations as well, i.e.\ one cannot ensure that the respective {\em lowest} grand potential in a systematic sequence of reference systems converges to the exact grand potential.
Namely, the stationary point with the lowest SFT grand potential could be a trivial stationary point (like one associated with a very simple reference system only as $H'_{\rm A}$ or $H'_{\rm B}$ in Fig.\ \ref{fig:ref}, for example).
Such an approximation must be considered as poor since the exact conditional equation for the self-energy is projected onto a very low-dimensional space only.

Therefore, one has to construct a different strategy which necessarily approaches the exact solution when following up a systematic sequence of reference systems.
Clearly, this can only be achieved if the following rule is obeyed:
{\em A non-trivial stationary point is always preferred as compared to a trivial one (R0).}
A non-trivial stationary point at a certain level of approximation, i.e.\ for a given reference system becomes a trivial stationary point on the next level, i.e.\ in the context of a ``new'' reference system that couples at least two different units of the ``old'' reference system.
Hence, by construction, the rule R0 implies that the exact result is approached for a systematic series of reference systems.

%*******************************************************************************
\begin{figure}[t]
\centerline{\includegraphics[width=0.94\textwidth]{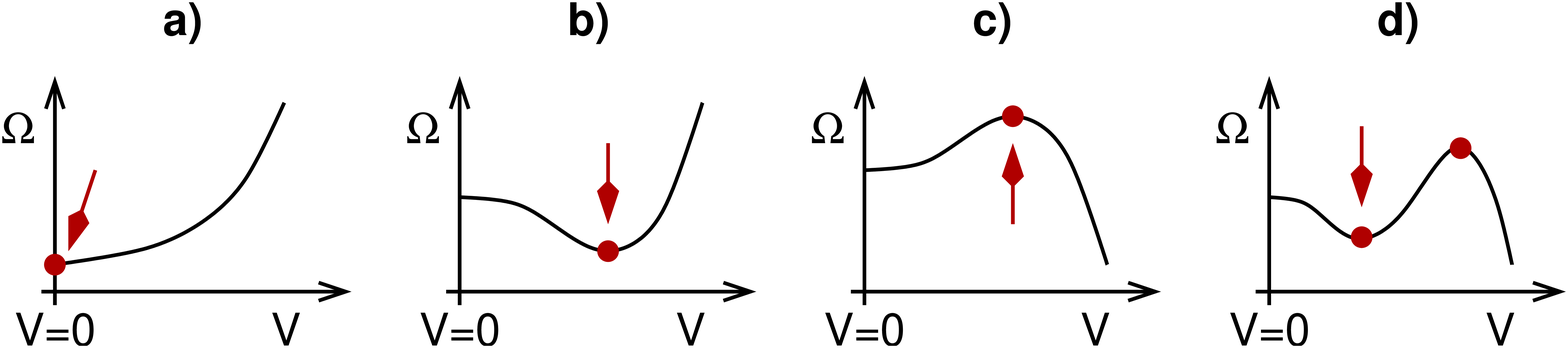}}
\caption{
Possible trends of the SFT grand potential $\Omega$ as a function of a variational parameter $V$ coupling two subsystems of a reference system. 
$V=0$ corresponds to the decoupled case and must always represent a ``trivial'' stationary point.
Circles show the stationary points to be considered.
The point $V=0$ has to be disregarded in all cases except for a).
The arrow marks the respective optimum stationary point according to the rules discussed in the text.
}
\label{fig:om}
\end{figure}
%*******************************************************************************

Following the rule (R0), however, may lead to inconsistent thermodynamic interpretations in case of a trivial stationary point with a lower grand potential as a non-trivial one.
To avoid this, R0 has to be replaced by:
{\em Trivial stationary points must be disregarded completely unless there is no non-trivial one (R1).}
This automatically ensures that there is at least one stationary point for any reference system, i.e.\ at any approximation level.

To maintain a thermodynamically consistent picture in case that there are more than a single {\em non-trivial} stationary points, we finally postulate:
{\em Among two non-trivial stationary points for the same reference system, the one with lower grand potential has to be preferred (R2).}

The rules are illustrated by Fig.\ \ref{fig:om}.
Note that the grand potential away from a stationary point does not have a direct physical interpretation.
Hence, there is no reason to interpret the solution corresponding to the maximum in Fig.\ \ref{fig:om}, c) as ``locally unstable''.
The results of Ref.\ \cite{Pot03b} (see Figs.\ 2 and 4 therein) nicely demonstrate that with the suggested strategy (R1, R2) one can consistently describe continuous as well as discontinuous phase transitions.

The rules R1 and R2 are unambiguously prescribed by the general demands for systematic improvement and for thermodynamic consistency.
There is no acceptable alternative to this strategy.
The strategy reduces to the standard strategy (always taking the solution with lowest grand potential) in case of the Ritz variational principle because here a non-trivial stationary point does always have a lower grand potential as compared to a trivial one.

There are also some consequences of the strategy which might be considered as disadvantageous but must be tolerated:
(i) For a sequence of stationary points that are determined by R1 and R2 from a systematic sequence of reference systems, the convergence to the corresponding SFT grand potentials is not necessarily monotonous.
(ii) Given two different approximations specified by two different reference systems, it is not possible to decide which one should be regarded as superior unless both reference systems belong to the same systematic sequence of reference 
systems.
In Fig.\ \ref{fig:ref}, one has $H'_{\rm A} < H'_{\rm B} < H'_{\rm C} < H'_{\rm D}$ where ``$<$'' stands for ``is inferior compared to''.
Furthermore, $H'_{\rm E} < H'_{\rm F}$ and $H'_{\rm A} < H'_{\rm E}$ but there is no relation between $H'_{\rm B}$ and $H'_{\rm E}$, for example.

\section{Bath sites and dynamical mean-field theory}
\label{sec:9}

\subsection{Motivation and dynamical impurity approximation}

Within self-energy-functional theory, an approximation is specified by the choice of the reference system. 
The reference system must share the same interaction part with the original model and should be amenable to a (numerically) exact solution.
These requirements restrict the number of conceivable approximations.
So far we have considered a decoupling of degrees of freedom by partitioning a Hubbard-type lattice model into finite Hubbard clusters of $L_c$ sites each, which results in the variational cluster approximation (VCA).

Another element in constructing approximation is to {\em add} degrees of freedom. 
Since the interaction part has to be kept unchanged, the only possibility to do that consists in adding new uncorrelated sites (or ``orbitals''), i.e.\ sites where the Hubbard-$U$ vanishes.
These are called ``bath sites''. 
The coupling of bath sites to the correlated sites with finite $U$ in the reference system via a one-particle term in the Hamiltonian is called ``hybridization''.

%*******************************************************************************
\begin{figure}[t]
\includegraphics[width=85mm]{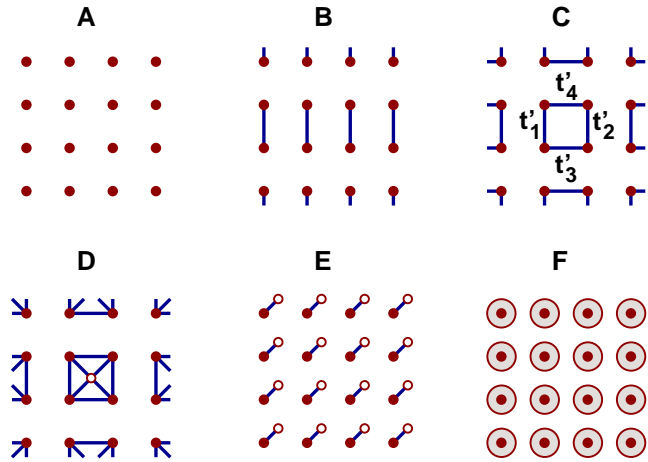}
\centering
\caption{
Different possible reference systems with the same interaction as the single-band Hubbard model on a square lattice.
Filled circles: correlated sites with $U$ as in the Hubbard model.
Open circles: uncorrelated ``bath'' sites with $U=0$.
Lines: nearest-neighbor hopping.
Big circles: continuous bath consisting of $L_{\rm b} = \infty$ bath sites.
Reference systems $H'_{\rm A}$, $H'_{\rm B}$, $H'_{\rm C}$ generate variational cluster approximations (VCA),
$H'_{\rm E}$ yields dynamical impurity approximation (DIA), $H'_{\rm F}$ the DMFT, and
$H'_{\rm D}$ an intermediate approximation (VCA with one additional bath site per cluster).
}
\label{fig:ref}
\end{figure}
%*******************************************************************************

Fig.\ \ref{fig:ref} shows different possibilities.
Reference system A yields a trial self-energy which is local $\Sigma_{ij\sigma}(\omega) = \delta_{ij} \Sigma(\omega)$ and has the same pole structure as the self-energy of the atomic limit of the Hubbard model. 
This results in a variant of the Hubbard-I approximation \cite{Hub63}. 
Reference systems B and C generate variational cluster approximations.
In reference system D an additional bath site is added to the finite cluster.
Reference system E generates a local self-energy again but, as compared to A, allows to treat more variational parameters, namely the on-site energies of the correlated and of the bath site and the hybridization between them. 
We call the resulting approximation a ``dynamical impurity approximation'' (DIA) with $L_b=1$. 

The DIA is a mean-field approximation since the self-energy is local which indicates that non-local two-particle correlations, e.g.\ spin-spin correlations, do not feed back to the one-particle Green function. 
It is, however, quite different from static mean-field (Hartree-Fock) theory since even on the $L_b=1$ level it includes retardation effects that result from processes $\propto V^2$ where the electron hops from the correlated to the bath site and back. 
This improves, as compared to the Hubbard-I approximation, the frequency dependence of the self-energy, i.e.\ the description of the temporal quantum fluctuations. 
The two-site ($L_b=1$) DIA is the most simple approximation which is non-perturbative, conserving, thermodynamically consistent and which respects the Luttinger sum rule (see Ref.\ \cite{OBP07}).
Besides the ``atomic'' physics that leads to the formation of the Hubbard bands, it also includes in the most simple form the possibility to form a local singlet, i.e.\ to screen the local magnetic moment on the correlated site by coupling to the local moment at the bath site.
The correct Kondo scale is missed, of course.
Since the two-site DIA is computationally extremely cheap, is has been employed frequently in the past, in particular to study the physics of the Mott metal-insulator transition \cite{Pot03a,Pot03b,KMOH04,Poz04,IKSK05a,IKSK05b,EKPV07,OBP07}.

\subsection{Relation to dynamical mean-field theory}

Starting from E and adding more and more bath sites to improve the description of temporal fluctuations, one ends up with reference system F where a continuum of bath sites ($L_b=\infty$) is attached to each of the disconnected correlated sites. 
This generates the ``optimal'' DIA.

To characterize this approximation, we consider the SFT grand potential given by \refeq{ocalc} and analyze the stationarity condition \refeq{stat}:
Calculating the derivative with respect to $\ff t'$, we get the general SFT Euler equation:
\be
   T \sum_{n} \sum_{\alpha\beta}
   \left( 
   \frac{1}{{\bf G}_{\ff t,0}^{-1}(i\omega_n) - {\ff \Sigma}_{\ff t',\ff U}(i\omega_n)} 
   -  {\bf G}_{\ff t',\ff U}(i\omega_n) \right)_{\beta \alpha} 
   \frac{\partial \Sigma_{\ff t',\ff U, \alpha\beta}(i\omega_n)}
        {\partial {{\bf t}'}}
   = 0  \: .
\nonumber \\   
\labeq{euler}
\ee
For the physical self-energy $\ff \Sigma_{\ff t,\ff U}$ of the original system $H_{\ff t,\ff U}$, the equation was fulfilled since the bracket would be zero.
Vice versa, since ${\ff G}_{\ff t',\ff U} = \widehat{\ff G}_{\ff U}[{\ff \Sigma}_{\ff t',\ff U}]$, the physical self-energy of $H_{\ff t,\ff U}$ is determined by the condition that the bracket be zero. 
Hence, one can consider the SFT Euler equation to be obtained from the {\em exact} conditional equation for the ``vector'' ${\ff \Sigma}$ in the self-energy space ${\cal S}_{\ff U}$ through {\em projection} onto the hypersurface of 
${\bf t}'$ representable trial self-energies ${\cal S}'_{\ff U}$ by taking the scalar product with vectors 
${\partial \Sigma_{\ff t',\ff U, \alpha\beta}(i\omega_n)}/{\partial {{\bf t}'}}$ tangential to the hypersurface.

Consider now the Hubbard model in particular and the trial self-energies generated by reference system F. 
Actually, F is a set of disconnected single-impurity Anderson models (SIAM).
Assuming translational symmetry, these impurity models are identical replicas.
The self-energy of the SIAM is non-zero on the correlated (``impurity'') site only.
Hence, the trial self-energies are local and site-independent, i.e.\ $\Sigma_{ij\sigma}(i\omega_n) = \delta_{ij} \Sigma(i\omega_n)$, and thus \refeq{euler} reads:
\be
   T \sum_{n} \sum_{i\sigma}
   \left( 
   \frac{1}{{\bf G}_{0}^{-1}(i\omega_n) - {\ff \Sigma}(i\omega_n)} - {\bf G}'(i\omega_n) 
   \right)_{ii\sigma} 
   \frac{\partial \Sigma_{ii\sigma}(i\omega_n)}
   {\partial {{\bf t}'}}
   = 0  \: ,
\labeq{euler1}
\ee
where the notation has been somewhat simplified.
The Euler equation would be solved if one-particle parameters of the SIAM and therewith an impurity self-energy can be found that, when inserted into the Dyson equation of the Hubbard model, yields a Green's function, the {\em local} element of which is equal to the impurity Green's function $\ff G'$. 
Namely, the bracket in \refeq{euler1}, i.e.\ the local (diagonal) elements of the bracket in \refeq{euler}, vanishes. Since the ``projector'' ${\partial \Sigma_{ii\sigma}(i\omega_n)}/{\partial {{\bf t}'}}$ is local, this is sufficient.

This way of solving the Euler equation, however, is just the prescription to obtain the self-energy within dynamical mean-field theory (DMFT) \cite{GKKR96}, and setting the local elements of the bracket to zero is just the self-consistency equation of DMFT.
We therefore see that reference system F generates the DMFT.
It is remarkable that with the VCA and the DMFT quite different approximation can be obtained in one and the same theoretical framework.

Note that for any finite $L_b$, as e.g.\ with reference system E, it is impossible to satisfy the DMFT self-consistency equation exactly since the impurity Green's function $\ff G'$ has a finite number of poles while the lattice Green's function $({1}/{{\bf G}_{0}^{-1}(i\omega_n) - {\ff \Sigma}(i\omega_n)})_{ii\sigma}$ has branch cuts.
Nevertheless, with the ``help'' of the projector, it is easily possible to find a stationary point $\ff t_{\rm opt}$ of the self-energy functional and to satisfy \refeq{euler1}.

%*******************************************************************************
\begin{figure}[t]
\includegraphics[width=90mm]{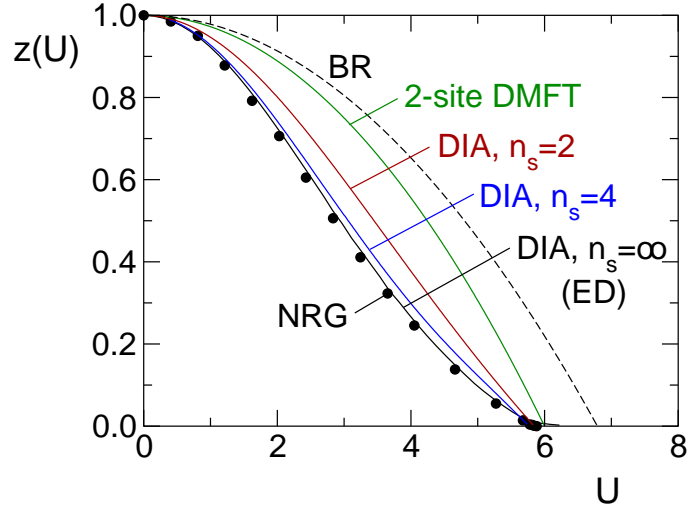}
\centering
\caption{
Quasiparticle weight $z$ as a function of $U$ for the half-filled Hubbard model at zero temperature as obtained by different approximations with local self-energy.
Here the lattice structure enters the theory via the non-interacting density of states $\rho_0(x)$ only. 
$\rho_0(x)$ is assumed to be semielliptic with band width $W=4$.
BR: Brinkman-Rice (Gutzwiller) result \cite{BR70}.
2S-DMFT: Two-site dynamical mean-field theory \cite{Pot01a}.
DIA: Dynamical impurity approximation with $n_s = 1+L_b = 2,4$ local degrees of freedom \cite{Pot03b}.
DIA, $n_s=\infty$: Full DMFT, evaluated using the DMFT-ED scheme with $n_s=8$ \cite{Pot03a}.
NRG: Full DMFT, evaluated using the numerical renormalization-group approach \cite{Bul99}.
}
\label{fig:z}
\end{figure}
%*******************************************************************************

Conceptually, this is rather different from the DMFT exact-diagonalization (DMFT-ED) approach \cite{CK94} which also solves a SIAM with a finite $L_b$ but which approximates the DMFT self-consistency condition. 
This means that within DMFT-ED an additional {\em ad hoc} prescription to necessary which, opposed to the DIA, will violate thermodynamical consistency.
However, an algorithmic implementation via a self-consistency cycle to solve the Euler equation is simpler within the DMFT-ED as compared to the DIA \cite{Poz04,EKPV07}. 
It has therefore been suggested to guide improved approximations to the self-consistency condition within DMFT-ED by the DIA \cite{Sen10}.

The convergence of results obtained by the DIA to those of full DMFT with increasing number of bath sites $L_b$ is usually fast.
With respect to the $U$-$T$ phase diagram for Mott transition in the single-band half-filled Hubbard model, $L_b =3$ is sufficient to get almost quantitative agreement with DMFT-QMC results, for example \cite{Poz04}.

This is also demonstrated with Fig.\ \ref{fig:z} which shows the quasiparticle weight,
\be
  z= \frac{1}{1 - d \Sigma(\omega=0) / d \omega} \: ,
\ee
as a function of $U$ for the Hubbard model with semielliptic free density of states (band width $W=4$) at half-filling and zero temperature. 
The Mott transition is marked by a divergence of the effective mass or, equivalently, by $z=0$ as $U$ approaches the critical value from below. 
It is remarkable that already the $n_s=1+L_b=4$-site DIA but also the $n_s=2$-site DIA (which requires the diagonalization of a dimer model only) yields $z(U)$ in an almost quantitative agreement with the full DMFT. 
The $n_s=2$-site-DIA may also be compared with the ``linearized'' or ``two-site'' DMFT \cite{BP00,Pot01a} where the Hubbard model is mapped onto the two-site single-impurity Anderson model by means a strongly simplified self-consistency condition. 
As compared to the two-site DMFT, the self-energy-functional approach not only represents a clear conceptual improvement but also improves the actual results for $z(U)$ and for $U_c$ (see Fig.\ \ref{fig:z} and Ref.\ \cite{Pot03a}).
Any arbitrariness in the method to find the parameters of the impurity model is avoided completely and consistent results are obtained for any finite $n_s$.

\subsection{Real-space dynamical mean-field theory}

Real-space DMFT \cite{PN99c,PN99a,PN99b,PN99d,HCR08,STT+08} with a local but site-dependent self-energy $\Sigma_{ij\sigma}(i\omega_n) = \delta_{ij} \Sigma_i(i\omega_n)$ is obtained from reference system F if the one-particle parameters of the single-impurity Anderson models at different sites are allowed to be different.
This becomes relevant for spatially inhomogeneous systems like semi-infinite lattices where surface effects can be studied or thin films or multilayers which are confined in one spatial direction or systems of fermions enclosed in an inhomogeneous external potential like gases of ultacold fermions in a confining harmonic potential.

%*******************************************************************************
\begin{figure}[b]
\includegraphics[width=65mm]{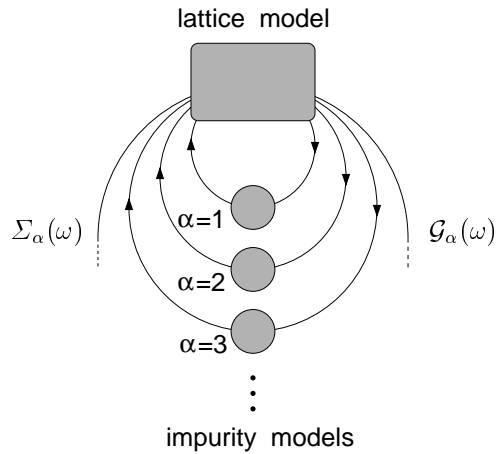}
\centering
\caption{
Generalized self-consistency cycle of the real-space dynamical mean-field theory, see text for discussion.
}
\label{fig:rdmft}
\end{figure}
%*******************************************************************************

Treating the different impurity models as inequivalent means to enlarge the space of trial self-energies.
Fig.\ \ref{fig:rdmft} shows the generalized self-consistency cycle of real-space DMFT.
The (inhomogeneous) lattice is partitioned into classes of sites which are equivalent due to residual spatial symmetries like, for example, lateral translational symmetry in case of a thin-film geometry.
A class of equivalent sites is denoted by $\alpha$. 
Within a given class the local self-energy is assumed to be site-independent while self-energies from different classes are allowed to be different.
For a completely disordered system without any spatial symmetry, the self-energy is fully site-dependent.
It this case $\alpha$ is a site index and the theory becomes equivalent with the so-called statistical DMFT \cite{DK97,DK98,BAF04,PB07}. 

As is sketched by Fig.\ \ref{fig:rdmft}, we start from a guess for the self-energy $\Sigma_\alpha(i\omega_n)$ for each class of sites. 
With this and with the free Green's function of the inhomogeneous lattice model, we get the interacting lattice Green's function $G_{ij\sigma}(i\omega_n)$ by solving Dyson's equation for the lattice model. 
Due to the reduced spatial symmetry, this is a matrix equation with a dimension given by the number of classes of inequivalent sites $\alpha$. 
\refeq{euler1} then provides us with one local DMFT self-consistency equation for each class which fixes the parameters of each of the inequivalent single-impurity Anderson models. 
The final and most important step in the cycle consists in solving the impurity models to get the self-energies $\Sigma_\alpha(i\omega_n)$ for all $\alpha$.
The different impurity models can be solved independently from each other in each step of the self-consistency cycle while the coupling between the different sites or classes of sites is provided by the Dyson equation of the lattice model.
 
\subsection{Cluster mean-field approximations}

As a mean-field approach, the DIA does not include the feedback of non-local two-particle correlations on the one-particle spectrum and on the thermodynamics. 
The DIA self-energy is local and takes into account temporal correlations only.
A straightforward idea to include short-range spatial correlations in addition, is to proceed to a reference system with $L_c>1$, i.e.\ system of disconnected finite clusters with $L_c$ sites each. 
The resulting approximation can be termed a cluster mean-field theory since despite the inclusion of short-range correlations, the approximation is still mean-field-like on length scales exceeding the cluster size. 

\begin{figure}[t]
\includegraphics[width=85mm]{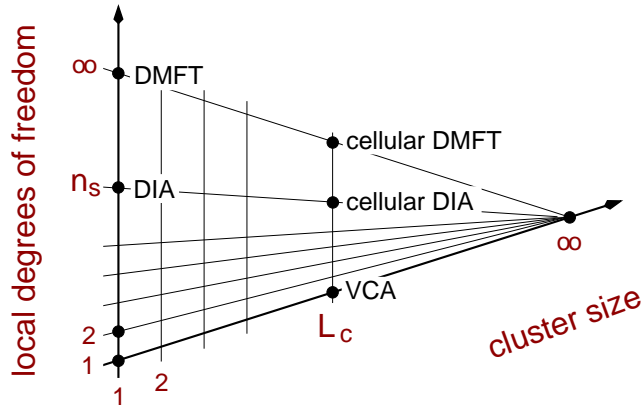}
\centering
\caption{
Dynamical impurity and cluster approximations generated by different reference systems within SFT.
$L_c$ is the number of correlated sites in the reference cluster.
$n_{\rm s}=1+L_b$ is the number of local degrees of freedom where $L_b$ denotes the number of additional bath sites attached to each of the $L_c$ correlated sites.
The variational cluster approximation (VCA) is obtained for finite clusters with $L_c>1$ but without bath sites $n_{\rm s}=1$. 
This generalizes the Hubbard-I-type approximation obtained for $L_c=1$.
The dynamical impurity approximation (DIA) is obtained for $n_{\rm s}>1$ but for a single correlated site $L_c=1$.
A continuum of bath sites, $n_{\rm s}=\infty$ generates DMFT ($L_c=1$) and cellular DMFT ($L_c>1$).
$L_c = \infty$, irrespective of the number of local degrees of freedom, corresponds to the exact solution.
}
\label{fig:approx}
\end{figure}

For an infinite number of bath degrees of freedom $L_b \to \infty$ attached to each of the $L_c >1$ correlated sites the cellular DMFT \cite{KSPB01,LK00} is recovered, see Fig.\ \ref{fig:approx}.
Considering a single-band Hubbard model again, this can be seen from the corresponding Euler equation: 
\be
   T \sum_{n} \sum'_{ij\sigma}
   \left( 
   \frac{1}{{\bf G}_{0}^{-1}(i\omega_n) - {\ff \Sigma}(i\omega_n)} - {\bf G}'(i\omega_n) 
   \right)_{ij\sigma} 
   \frac{\partial \Sigma_{ji\sigma}(i\omega_n)}
   {\partial {{\bf t}'}}
   = 0  \: ,
\labeq{euler2}
\ee
where the prime at the sum over the sites indicates that $i$ and $j$ must belong to the same cluster of the reference system.
Namely, $\Sigma_{ij\sigma}(i\omega_n) = 0$ and also the ``projector'' ${\partial \Sigma_{ji\sigma}(i\omega_n)}/{\partial {{\bf t}'}} = 0$ if $i$ and $j$ belong to different clusters.
This stationarity condition can be fulfilled if 
\be
   \left( 
   \frac{1}{{\bf G}_{0}^{-1}(i\omega_n) - {\ff \Sigma}(i\omega_n)} - {\bf G}'(i\omega_n) 
   \right)_{ij\sigma} 
   = 0  \: .
\labeq{cdmft}
\ee
Note that ${\bf G}'(i\omega_n)$ is a matrix which is labeled as 
${G}'_{ij,kl}(i\omega_n)$ where $i,j=1,...,L_c$ refer to the correlated sites in the cluster while $k,l = 1,...,L_bL_c$ to the bath sites. 
${G}'_{ij}(i\omega_n)$ are the elements of the cluster Green's function on the correlated sites. 
The condition \refeq{cdmft} is just the self-consistency condition of the C-DMFT.

As is illustrated in Fig.\ \ref{fig:approx}, the exact solution can be obtained with increasing cluster size $L_c\to\infty$ either from a sequence of reference systems with a continuous bath $n_{\rm s}=1+L_b=\infty$, corresponding to C-DMFT, or from a sequence with $n_{\rm s}=1$, corresponding to VCA, or with a finite, small number of bath sites (``cellular DIA''). 
Systematic studies of the one-dimensional Hubbard model \cite{PAD03,BHP08} have shown that the energy gain which is obtained by attaching a bath site is lower than the gain obtained by increasing the cluster.
This suggests that the convergence to the exact solution could be faster on the ``VCA axis'' in Fig.\ \ref{fig:approx}. 
For a definite answer, however, more systematic studies, also in higher dimensions, are needed.

%*******************************************************************************
%*******************************************************************************

In any case bath sites help to get a smooth dependence of physical quantities when varying the electron density or the (physical) chemical potential.
The reason is that bath sites also serve as ``charge reservoirs'', i.e.\ during a $\mu$ scan the ground state of the reference cluster may stay in one and the same sector characterized by the conserved total particle number in the cluster while the particle number on the correlated sites and the approximate particle number in the original lattice model evolve continuously \cite{BHP08,BP10}.
This is achieved by a $\mu$-dependent charge transfer between correlated and bath sites.
In addition, (at least) a single bath site per correlated site in a finite reference cluster is also advantageous to include the interplay between local (Kondo-type) and non-local (antiferromagnetic) singlet formation. 
This has been recognized to be important in studies of the Mott transition \cite{BKS+09} in the two-dimensional and of ferromagnetic order in one-dimensional systems \cite{BP10}.
For studies of spontaneous U(1) symmetry breaking, e.g.\ $d$-wave superconductivity in the two-dimensional Hubbard model \cite{SLMT05,AA05,AAPH06a,AAPH06b}, doping dependencies can be investigated without bath sites due to mixing of cluster states with different particle numbers.

\begin{figure}[t]
\centering
\includegraphics[width=0.65\columnwidth]{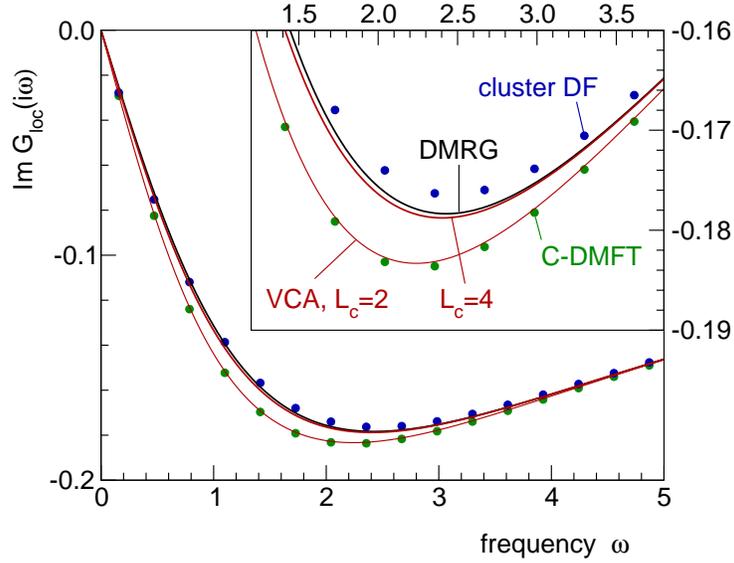}
\caption{(taken from Ref.\ \cite{BHP08}) 
Local Green's function on the Matsubara frequencies, $G_{ii\sigma}(i\omega_n)$, for $U=6$ using the $L_c=2$, $n_s=3$ and the $L_c=4$, $n_s=3$ reference systems (red solid lines).
Zero-temperature results for the one-dimensional Hubbard model at half-filling with nearest-neighbor hopping set to $t=1$ are compared to corresponding dynamical DMRG data (black solid line), $L_c=2$ cellular DMFT at finite but low temperature ($\beta=20$) and $L_c=2$ cluster dual-fermion theory (cluster DF, $\beta=20$) from from Ref.\ \cite{HBR+07}.
}
\label{fig:green}
\end{figure}

To benchmark the cluster mean-field approximations that can be constructed from the dynamical variational principle,  it is instructive to compare with numerically exact data.
Essentially exact results from the dynamical density-matrix renormalization-group approach (DMRG) are available for the one-dimensional Hubbard model at zero temperature (see Ref.\ \cite{HBR+07}). 
A one-dimensional model with strong spatial correlations is actually the least favorable limit to start off with a cluster approach.
Fig.\ \ref{fig:green} shows the result for the local one-particle Green's function on the Matsubara frequencies as obtained from the variational cluster approximation with $L_c=2$ correlated sites and with 2 bath sites per correlated site in addition ($n_s=3$). 
While the overall trend is reproduced well, there are also apparent discrepancies at intermediate frequencies. 
The comparison with the cellular DMFT result using weak-coupling continuous-time QMC at finite temperature ($1/T = \beta = 20$), however, shows that there is hardly any improvement when a continuum of bath sites is attached ($n_s=\infty$), i.e.\ the main effect is captured with a very small number of local degrees of freedom $n_s$ already.
The cluster dual-fermion method \cite{HBR+07} on the other hand represents an approach that goes beyond the cluster mean-field theory by including diagrams contributing to the inter-cluster elements of the self-energy. 
For $L_c=2$ this is clearly superior as compared to the VCA for $L_c=2$ (and $\beta=20$) and already quite close to the DMRG data.
It should be noted, however, that the VCA calculations are computationally much cheaper.
Using a larger cluster, i.e.\ an $L_c=4$ reference system with bath sites ($n_s=3$) attached to the cluster edges only, the VCA Green's function can hardly be distinguished from the dynamical DMRG result.

\subsection{Translation symmetry}

For any cluster approximation formulated in real space there is an apparent problem: 
Due to the construction of the reference system as a set of decoupled clusters, the trial self-energies do not preserve the translational symmetries of the original lattice. 
Trivially, this also holds if periodic boundary conditions are imposed for the individual cluster.
Transformation of the original problem to reciprocal space does not solve the problem either since this also means to transform a local Hubbard-type interaction into a non-local interaction part which basically couples all $\ff k$ points.

There are different ideas to overcome this problem.
We introduce a ``periodizing'' universal functional 
\be
  \widehat{T}[\ff \Sigma]_{ij} = \frac{1}{L} \sum_{i'j'} \delta_{i-j,i'-j'} \Sigma_{i'j'}
\ee
which maps any trial self-energy onto translationally invariant one.
In reciprocal space this corresponds to the substitution $\Sigma_{\ff k,\ff k'} \to \widehat{T}[\ff \Sigma]_{\ff k,\ff k'} = \delta_{\ff k,\ff k'} \Sigma_{\ff k}$.
Using this, we replace the self-energy functional of \refeq{sfp} by
\be
  \widehat{\Omega}^{(1)}_{\ff t', \ff U}[\ff \Sigma] = 
  \Tr \ln \frac{1}{\ff G_{\ff t',0}^{-1} - {\widehat \ff T}[\ff \Sigma]}
  + \widehat{F}_{\ff U}[\ff \Sigma] \: ,
\labeq{sfpnew}
\ee
as suggested in Ref.\ \cite{KD05}.
This new functional is different from the original one. 
However, as the physical self-energy is supposed to be translational invariant, it is a stationary point of both, the original and the modified functional.
This means that the modified functional can likewise be used as a starting point to construct approximations.
It turns out (see Ref.\ \cite{PB07} for an analogous discussion in case of disorder) that for a reference system with $L_c>1$ and $n_{\rm s}=\infty$, the corresponding Euler equation reduces to the self-consistency equation of the so-called periodized cellular DMFT (PC-DMFT) \cite{BPK04}.
The same modified functional can also be used to construct a periodized VCA, for example.

While the main idea to recover the PC-DMFT is to modify the form of the self-energy functional, the dynamical cluster approximation (DCA) \cite{HTZ+98,MJPK00,HMJK00} is obtained with the original functional but a modified hopping term in the Hamiltonian of the original system. 
We replace $\ff t \to \widetilde{\ff t}$ and consider the functional $\Omega_{\widetilde{\ff t}, \ff U}[\ff \Sigma]$.
To ensure that the resulting approximations systematically approach the exact solution for cluster size $L_c \to \infty$, the replacement $\ff t \to \widetilde{\ff t}$ must be controlled by $L_c$, i.e.\ it must be exact up to irrelevant boundary terms in the infinite-cluster limit.
This is the case for 
\be
  \widetilde{\ff t} = (\ff V \ff W) \ff U^\dagger \ff t \ff U (\ff V\ff W)^\dagger \: ,
\ee
where $\ff U$, $\ff V$, and $\ff W$ are unitary transformations of the one-particle basis. 
$\ff U$ is the Fourier transformation with respect to the original lattice consisting of $L$ sites ($L\times L$ unitary matrix).
$\ff W$ is the Fourier transformation on the cluster ($L_c \times L_c$), and 
$\ff V$ the Fourier transformation with respect to the superlattice consisting of $L/L_c$ supersites given by the clusters ($L/L_c \times L/L_c$).
The important point is that for any finite $L_c$ the combined transformation $\ff V \ff W = \ff W \ff V \ne \ff U$, while this becomes irrelevant in the limit $L_c\to \infty$.
The detailed calculation (see Ref.\ \cite{PB07} for the analogous disorder case) shows that the DCA is recovered for a reference system with with $L_c>1$ and $n_{\rm s}=\infty$, if periodic boundary conditions are imposed for the cluster.
The same modified construction can also be used to a get simplified DCA-type approximation without bath sites, for example.
This ``simplified DCA'' is related to the periodized VCA in the same way as the DCA is related to the PC-DMFT.
The simplified DCA would represent a variational generalization of a non-self-consistent approximation (``periodic CPT'') introduced recently \cite{MT06}.

\section{Concluding discussion}
\label{sec:10}

The physical properties of strongly correlated electron systems are often dominated by a complex interplay of a macroscopically large number of degrees of freedom that cannot be described by means of brute-force numerical methods:
The exact-diagonalization or Lanczos technique \cite{LG93,Fre00} is restricted to systems with moderately large Hilbert space dimensions only, the numerical renormalization-group approach \cite{Wil75} as well as the density-matrix renormalization group \cite{Whi92,Sch10} are designed for impurity or one-dimensional models only, and the quantum Monte-Carlo technique \cite{BSS81,HF86,GML+11} suffers from the famous sign problem for most of the two- or higher-dimensional lattice fermion models.

A completely different approach to strongly correlated systems is offered by mean-field theory. 
Common to the huge number of different versions of mean-field concepts is that certain correlations between the degrees of freedom are neglected to make the problem tractable analytically or accessible to one of the aforementioned numerical methods.
Hartree-Fock theory is one of the most clear and also one of the most frequently employed mean-field approaches. 
For an interacting $N$-electron system, the Hartree-Fock approximation decouples the dynamics of a particular electron from the rest of the system by setting up an effective one-particle Schr\"odinger equation which includes an effective one-electron potential which is self-consistently produced by the remaining $N-1$ electrons.

We have seen that a very elegant way to derive the Hartree-Fock theory for a system with a rather general second-quantized Hamiltonian is to employ a generalized version of the Ritz variational principle $\delta \Omega_{\ff t,\ff U}[\rho]=0$. 
To get an ansatz for the density operator $\rho$, it has turned out to be convenient in general to assume $\rho$ to be given as the exact density operator $\rho_{\ff t',\ff U'}$ of a certain reference system with different one-particle and interaction parameters $\ff t'$ and $\ff U'$. 
Hartree-Fock theory is then obtained with $\ff U'=0$ but $\ff t'$ arbitrary, i.e.\ the trial density operator is parameterized by $\ff t'$ and the grand potential is optimized by varying $\rho_{\ff t',0}$ through variation of $\ff t'$.

The essential point why this concept is working at all, consists in the fact that due to Wick's theorem the density-operator functional $\Omega_{\ff t,\ff U}[\rho]$ can be evaluated exactly on the subspace of trial density operators while this is generally impossible in practice for an arbitrary $\rho$.
This is the reason why other approaches based on the static variational principle, like the Gutzwiller approach \cite{Gut63,Geb06}, for example, have to employ additional approximations or, like the variation of matrix-product states \cite{Sch10}, need additional heavy numerical tools and are restricted to certain limits, as e.g.\ to one-dimensional models.

Loosely speaking, the dynamical variational principle states that the grand potential of an interacting electron system is stationary at the physical Green's function or the physical self-energy.
It actually goes without saying that this crucially depends on the functional form of $\widehat{\Omega}_{\ff t,\ff U}[\ff \Sigma]$.
In particular, to formulate a dynamical variational principle, it is inevitable to involve the Luttinger-Ward functional in the one or other way or similar functionals such as its Legendre transform.
The Luttinger-Ward functional can either be defined order by order in diagrammatic perturbation theory or by means of the fermionic path integral. 
Likewise, the dynamical variational principle can be formulated in different equivalent variants, namely as stationarity of the grand potential as a functional of the Green's function, of the self-energy or as a functional of both. 
A slight conceptual advantage of the self-energy-functional approach is that dynamical mean-field theory can be obtained by merely restricting the domain of the functional, i.e.\ in the same way as Hartree-Fock theory is obtained from the static Ritz principle.

The dynamical variational principle is in fact known since long \cite{LW60} but has not been employed to construct non-perturbative approximations until recently \cite{Pot03a}.
In fact, a direct evaluation of the Luttinger-Ward functional for a given Green's function is a hopeless task since this implies the full summation of skeleton diagrams of arbitrary order.
The essential point is that to make use of the principle one can proceed analogously to the construction of static mean-field theory from the Ritz principle and choose a reference system that spans a certain space of trial self-energies such that the self-energy functional can be evaluated exactly there.
This is in fact possible if reference systems are considered that share with the original system the same interaction part but are accessible to a numerically exact solution.
The underlying reason is that the form of the Luttinger-Ward functional is exactly the same in both cases as this parametrically depends on the vertices only and that therewith the functional can be evaluated for Green's functions of the reference system at least.

For both, the static and the dynamical variational principle, type-III approximations, i.e.\ approximations that merely result from a restriction of the domain of the functional as specified by the choice of a reference system, can be constructed in a very similar way -- at least formally. 
In practice, however, the resulting approximations are very different. 
For the static variational principle, the only meaningful reference system that still allows for an exact evaluation is given by $\ff U'=0$ but $\ff t'$ arbitrary. 
This implies that one is looking for the optimal non-interacting density operator for an interacting system and what is found is Hartree-Fock theory. 
For the dynamical variational principle, on the other hand, the only meaningful reference systems are those where $\ff U'=\ff U$ and $\ff t'$ such that the reference system's Hamiltonian describes a set of spatially decoupled subsystems with small Hilbert-space dimension respectively.
The prototypical example is the variational cluster approximation or, if local variational degrees of freedom are added, the (cellular) dynamical mean-field theory.
Hence, while static theory produces non-local but also effectively non-interacting density operators, the dynamical approach leads to local but non-perturbative self-energies.

A different way of stating this is to say that dynamical objects, the self-energy or the Green's function, are necessarily to be considered as the basic quantities of the theory if one aims at an embedding approach, i.e.\ at an approach where one likes to expand around a local or cluster model and where such an expansion is sensible as the dominating correlations are local.
Note that the same concept fails if one tries to base this on the static variational principle: 
Choosing with $\ff t'$ and $\ff U$ a reference system of decoupled sites or clusters with the same interaction part as the original (Hubbard-type) model, spans a space of {\em density operators} which factorize spatially. 
This implies, however, that {\em any} spatial correlation function vanishes completely. 
This would represent a too strong approximation that cannot be tolerated in any circumstance.

It is interesting to note that only the dynamical variational principle allows for a systematic improvement of approximations, namely simply by enlarging the cluster size in the reference system. 
This ensures that, at least in principle, one can approach the exact solution. 
Opposed to this, there is no way of improving on the Hartree-Fock approximation without giving up the concept of type-III approximations.
For any finite $\ff U'$ in the reference system, Wick's theorem is no longer applicable.

Another quite important difference between the static and the dynamical approach shows up for systems with non-local interaction, i.e.\ for models beyond the simple Hubbard model.
While nothing actually happens to the formal construction of static mean-field theory, a dynamical mean-field theory can no longer be formulated from the dynamical variational principle:
There is simply no reference system at all that can be solved numerically exact: 
As we have seen, to circumvent the practically impossible direct evaluation of the Luttinger-Ward functional, $\ff U' =\ff U$ must be assumed, and thereby the reference system's degrees of freedom remain coupled.
An interesting possible way out which is worth to be explored in the future, is to set up a generalized dynamical variational principle where the $\ff U$-dependent but $\ff t$-independent Luttinger-Ward functional is replaced by a truly universal, i.e.\ $\ff t$ {\em and} $\ff U$-independent functional. 
This would allow to consider reference systems with arbitrary parameters $\ff t'$ and $\ff U'$ and with $\ff U'\ne \ff U$ in particular. 

Finally, it should be stressed that despite all the close analogies there is no known strict correspondence between the static and the dynamical variational principle and no correspondence between the approximations constructed within the different frameworks. 
For example, the density operator of the system's thermal state within dynamical mean-field theory cannot be specified explicitly. 
Within a dynamically constructed approximation, such as DMFT, the approximate quantum state of the original lattice system that corresponds to the DMFT Green's function or self-energy is not known.
Vice versa, static mean-field theory cannot be derived from the dynamical variational principle as a type-III approximation: Choosing a reference system with $\ff U'=0$ implies to shrink the space of trial self-energies to a single point $\ff \Sigma = 0$ which is different from the Hartree-Fock self-energy.
The missing one-to-one correspondence between static and dynamical approaches also reflects itself in the fact 
that neither the self-energy functional nor any of the suggested variants are generally concave (or convex). 
Stationary points are usually saddle points instead of local or global extrema. 
Accordingly, dynamical approximations cannot be shown to provide strict upper bounds to the true grand potential of the system. 
This is opposed to {\em any} static approximation, including even the crudest ones. 

\begin{theacknowledgments}
These lecture notes heavily build on material to be published as a contribution to a book on {\em Strongly correlated systems: methods and techniques} \cite{Pot11a}. 
The opportunity to lecture at the XV Training Course in the Physics of Strongly Correlated Systems (Vietri sul Mare, Salerno, Italy in October 2010) gave me reason to extend and sharpen the topic and particularly to show up the close relation and the differences to concepts of static mean-field theory.
I would like to thank the organizers of the Training Course, Professors Adolfo Avella and Ferdinando Mancini, for the invitation.
I also would like to thank
M. Aichhorn,
F.F. Assaad,
E. Arrigoni,
M. Balzer,
R. Bulla,
C. Dahnken,
R. Eder,
W. Hanke,
A. Hewson,
M. Jarrell,
M. Kollar,
G. Kotliar,
A.I. Lichtenstein,
A.J. Millis,
W. Nolting,
D. Senechal,
A.-M.S. Tremblay, 
D. Vollhardt
for cooperations and many helpful discussions over the recent years.
The work was supported in part by the Deutsche Forschungsgemeinschaft through SFB 668 (project A14) and FOR 1346 (project P1).
\end{theacknowledgments}

\bibliographystyle{aipproc}   % if natbib is available


\end{document}